\documentclass[twocolumn, prl, preprintnumbers,amsmath,amssymb, showkeys]{revtex4}
\usepackage{graphicx}
\usepackage{dcolumn}
\usepackage{bm}
\usepackage{booktabs}
\usepackage{color}
\usepackage{ifthen}
\usepackage{hyperref}
\usepackage{url}
\graphicspath{{./figures/}}
\keywords{compact polymers, distance distribution, chromatin folding}
\begin{document}

\title{Conformational properties of compact polymers\footnote{Copyright (2009) American Institute of Physics. This article may be downloaded for personal use only. Any other use requires prior permission of the author and the American Institute of Physics.

Article published in: \textit{J. Chem. Phys.} \textbf{130}, 174901 (2009); DOI:10.1063/1.3126651;\\ URL \url{http://link.aip.org/link/?JCPSA6/130/174901/1}}}

\author{Manfred Bohn}
\email{bohn@tphys.uni-heidelberg.de}
\author{Dieter W. Heermann}
\affiliation{Institute of Theoretical Physics, University of Heidelberg, Philosophenweg 19, D-69120 Heidelberg, Germany}
\date{\today}
\begin{abstract}
Monte Carlo simulations of coarse-grained polymers provide a useful tool to deepen the understanding of conformational and statistical properties of polymers both in physical as well as in biological systems. In this study we sample compact conformations on a cubic $L\times L\times L$ lattice with different occupancy fractions by modifying a recently proposed algorithm. The system sizes studied extend up to $N=256\,000$ monomers, going well beyond the limits of older publications on compact polymers. We analyze several conformational properties of these polymers, including segment correlations and screening of excluded volume. Most importantly we propose a scaling law for the end-to-end distance distribution and analyze the moments of this distribution. It shows  universality with respect to different occupancy fractions, i.e. system densities. We further analyze the distance distribution between intrachain segments, which turns out to be of great importance for biological experiments. We apply these new findings to the problem of chromatin folding inside interphase nuclei and show that -- although chromatin is in a compacted state -- the classical theory of compact polymers does not explain recent experimental results. 
\end{abstract}

\maketitle
\section{Introduction}
Compact polymers are of significant interest from the viewpoint of polymer physics~\cite{Grosberg1994}. They describe the conformational behavior of polymers obeying excluded volume interactions which are brought into a bad solvent, meaning that the solvent molecules effectively tend to repell the chain molecules. This leads to the formation of compact arrangements of the monomers as the polymer tries to minimize the solvent chain interface. From a different viewpoint compact conformations also arise when there are attractive forces acting between the constituents of the polymer which are larger than the entropic forces trying to enhance the radius of gyration of the polymer. A polymer is said to be compact or globular if its characteristic size scales with $N^{1/3}$, resulting in a nearly uniform monomer concentration inside a globule of a certain radius scaling with $N^{1/3}$.  However, besides the pure physical interest in compact polymers, it is also of great importance to study such conformations from the biological point of view. Many polymers in living organisms tend to organize in a compact way, the prime example are proteins~\cite{Dill1995}. A lot of studies have been devoted to the problem of protein folding starting from the sampling of random compact conformations~\cite{Lau1989, Shakhnovich1993}. There is also evidence that chromatin inside the interphase nucleus is organized at least partly in a very compacted state~\cite{Munkel1998, Goetze2007, Mateos2009}. 

Here we want to deepen the understanding of the class of compact polymers with respect to its conformational and statistical properties using a Monte Carlo method. Monte Carlo simulations have proven to be a very important technique to study ensembles of polymer chains. Much progress has been achieved for example by using simple lattice models to study self-avoiding walks~\cite{Eizenberg1993, Madras1988, Rapaport1985, Domb1969}. 
Obviously, Monte Carlo techniques become unavoidable when analytical or exact enumeration methods fail to be feasible. Indeed, analytical methods become very rare when studying problems where excluded volume has to be taken into account, which is the case in most applications. Exact enumeration methods~\cite{Shakhnovich1990, Pande1996}, where the whole configuration space is sampled and exact averages can be calculated, are limited to very small system sizes which most often do not show the correct asymptotic behavior for large chains.

In the last few years several Monte Carlo algorithms have been proposed to study compact polymers based on the idea of Hamiltonian paths~\cite{Ramakrishnan1995, Mansfield2006, Lua2004}. A Hamiltonian path on some graph $G$ with set of vertices $\mathcal{V}$ and edges $\mathcal{E}$ is defined as a path which visits each vertex $V\in\mathcal{V}$ exactly once. Obviously, Hamiltonian paths studied on a cubic lattice are prime examples of maximally compact polymers where the number of nearest neighbor contacts is maximized. However, the exact enumeration of all possible conformations is not feasible as the computer resources needed grow exponentially with growing system size~\cite{Pande1996}. Thus such studies are limited to rather small system sizes which probably do not reflect the properties of compact polymers in the limit of large chain lengths $N$. Therefore, several Monte Carlo techniques have been developed to sample a representative ensemble of Hamiltonian paths on a cubic lattice. One important aspect of such an algorithm is that sampling of conformations is done in an unbiased way. Two algorithms have been shown to fail this test~\cite{Ramakrishnan1995, Lua2004}. Recently, Mansfield proposed an algorithm which is shown to produce unbiased samples to a high degree of certainty~\cite{Mansfield2006}.

The scope of this study is three-fold. First of all, we extend the algorithm proposed by Mansfield~\cite{Mansfield2006} in order to study not only maximally compact conformations where all lattice sites on a cubic lattice are occupied, but also conformations in less dense systems with density $\rho\neq1$. By using a highly parallel system we generate chains  of lengths much larger than the ones studied in previous publications. The largest system size for $\rho=1$ is $L=55$, the largest chain length studied is $N=256\,000$ for a density of $\rho=0.5$.  Secondly, we analyze several statistical and conformational properties of these compact conformations. Special interest is on the distance distribution between the end points of the chain as well as the distance distribution between smaller segments of the chain. Stunningly this quantity has not been analyzed in previous publications, although it is of severe importance for biological applications as the distributions can be compared directly to experimental data~\cite{Mateos2009}. Furthermore, we provide a comparison of compact polymers to a polymer melt of equal density, which has been suggested to behave similarly~\cite{Grosberg1994, Lua2004}. In a third step, we apply the results of our simulational study to recent experimental data concerning the folding of chromatin inside the interphase nucleus. We show that chromatin on a scale above 150 kb does not organize simply in a compact state as the behaviour of the mean square displacement with genomic distance might suggest~\cite{Mateos2009}, but shows important hallmarks of a disordered system as described by the random loop model~\cite{Bohn2007a, Mateos2009}.

\section{The Computational Method}
\subsection{Algorithm}
To create conformations of compact polymers on a cubic lattice we use a modified version of the algorithm Mansfield~\cite{Mansfield2006} proposed for sampling Hamiltonian paths. This algorithm is a Metropolis Monte Carlo technique and is one of the few known algorithms except exact enumeration methods which is unbiased, i.e. every allowed conformation is sampled with equal probability. While other proposed algorithms for sampling Hamiltonian paths have been shown to produce biased results~\cite{Ramakrishnan1995, Lua2004},   Mansfield proved in his paper~\cite{Mansfield2006} ergodicity for small lattices by exact enumeration and devised a method providing strong evidence that the algorithm is ergodic for even larger lattices.

Here we are not only interested in Hamiltonian paths, which only make up a subset of compact polymers but also in compact conformations within the finite temperature regime where $\rho \neq 1$. Therefore we have to allow vacancies on the simulation lattice. This is done by modifying the algorithm such that it also handles lattices where not all vertices are occupied by introducing a reptation step for the chain ends. 

Consider a cubic lattice of dimension $L\times L\times L$, each lattice side $\mathbf{r}=(x,y,z)$ is either occupied or unoccupied. The chain itself is stored in a list of lattice sites. The algorithm then works as follows:
\begin{enumerate}
 \item Randomly select one of the two ends of the chain, the coordinates denoted by $\mathbf{r_E}$.
 \item Randomly select one of the six neighbouring sites of  $\mathbf{r_E}$ on the cubic lattice, denoted by $\mathbf{r_N}$. 
 \item Test if $\mathbf{r_N}$ lies outside the lattice. If so, proceed with (5), otherwise proceed with (4). 
 \item Test if the lattice site $\mathbf{r_N}$ is occupied. If the lattice site is occupied and $\mathbf{r_E}$ lies at the head of the list, we reverse the part of the list lying above $\mathbf{r}_N$. If $\mathbf{r_E}$ lies at the tail of the list, we reverse the part of the list lying below $\mathbf{r_N}$. 
If the lattice site is unoccupied we do a reptation move, i.e. we append the position $\mathbf{r_N}$ to the head (if $\mathbf{r_E}$ is currently head) or to the tail (if $\mathbf{r_E}$ is currently tail) of the list and remove the other end of the list. 
 \item Take the new conformation (if or if not it has changed) as the current configuration. 
\end{enumerate}
This algorithm equals the one proposed by Mansfield for compact polymers with $\rho=1$, as in this case reptation steps are not possible. The only change applied to the algorithm is that we allow for a reptation step in (4) whenever the lattice site $\mathbf{r}_N$ is unoccupied. 

\subsection{Ergodicity and unbiased sampling}
The algorithm is a Monte Carlo technique, which means that we have to make sure that it produces unbiased samples. Unbiased means that a) the algorithm  samples the complete configuration space and that b) each configuration is sampled with equal probability. Unbiased sampling is ensured if the algorithm satisfies detailed balance
and ergodicity. Detailed balance is satisfied obviously by the algorithm. However, it is not clear a priori that ergodicity is satisfied. There is strong evidence that the algorithm satisfies ergodicity for $\rho=1$~\cite{Mansfield2006}, but we have to make sure that this also holds for our modified algorithm. The reptation algorithm on its own is non-ergodic~\cite{Binder2002} as a conformation can get trapped in a state where no more reptation moves are possible. However, this problem is resolved here: Whenever there is no free adjacent lattice site to a chain end, the algorithm performs a half-list reversal such that the chain's end moves to another lattice site. 
To test the ergodicity of the algorithm quantitatively, we performed an ergodicity test on a small $3\times3\times3$ lattice in the following manner: First we enumerate all possible walks of given length $N$ on the lattice by exact enumeration. Each of these conformations are assigned an equivalence class $A_i$. An equivalence class $A_i$ ($i=1, \ldots, M$) represents all walks which are equal after applying translational shift (for $\rho\neq1$), one of the 47  symmetry operations on the cubic lattice or a path reversal. It is obvious to subsume all conformations with rotational and translational symmetry into one equivalence class, as we are not interested in properties depending on the rotational state or absolute position in space. To include path reversal is necessary for the algorithm to be ergodic for $\rho=1$~\cite{Mansfield2006}. For example there are $51\,704$ equivalence classes for $N=27$ and $2\,750$ classes for $N=10$ on a $3\times3\times3$ lattice.

\begin{figure}
 \includegraphics[width=\hsize]{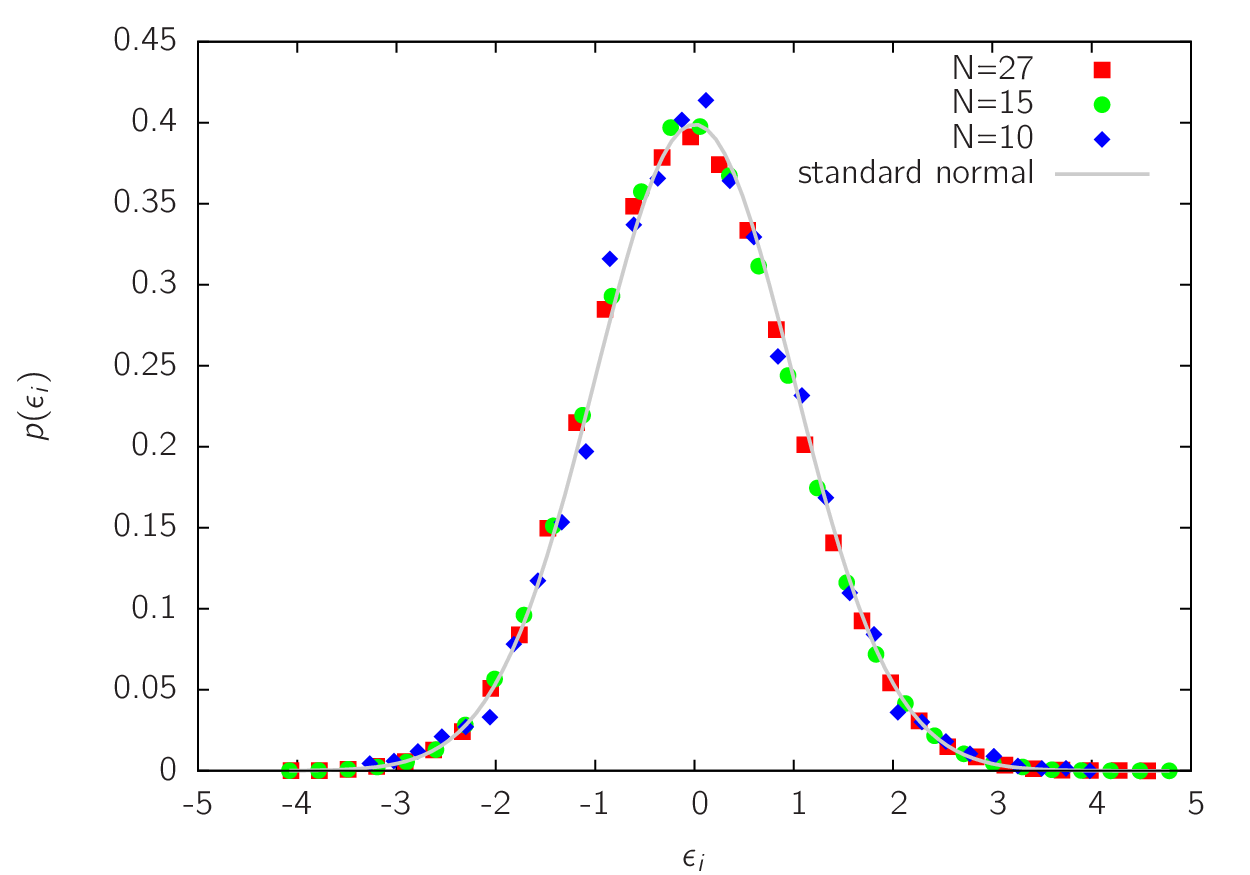}
 \caption{\label{fig:ergodicity} Ergodicity test of the algorithm for polymers on a $3\times3\times3$-lattice. Shown is the probability distribution of $\epsilon_i$, given in eq.~\eqref{eq:epsilon}. The distribution $p(\epsilon_i)$ is in good agreement with the standard normal distribution, which is evidence that the algorithm is ergodic.}
\end{figure}

Let $p_i$ be the probability for a conformation to fall in equivalence class $A_i$, determined by exact enumeration. If we then produce $K$ independent samples by the Monte Carlo algorithm, the probability that $k$ samples fall in equivalence class $A_i$ then is 
\[ P_i(k) = \binom{K}{k} p_i^k (1-p_i)^{K-k} \]
Let $q_i$ denote the relative abundance that from our  $K$ samples a randomly chosen one falls into equivalence class $A_i$. If the algorithm is unbiased, then the random variable
\begin{equation}\label{eq:epsilon}
 \epsilon_i = \frac{q_i-p_i}{\sqrt{\frac{p_i(1-p_i)}{K}}}
\end{equation}
is normal distributed with mean zero and variance equal to unity. Fig.~\ref{fig:ergodicity} shows that the $\epsilon_i$ are in a very good approximation normal distributed for different chain length $N$. Deviations from the normal distribution most probably are due to the fact that the $\epsilon_i$ are not independent. As there is no reason to believe that ergodicity is broken for larger lattice sizes, our modified algorithm most probably satisfies ergodicity. 

\subsection{Autocorrelation times}
\begin{figure}
 \includegraphics[width=\hsize]{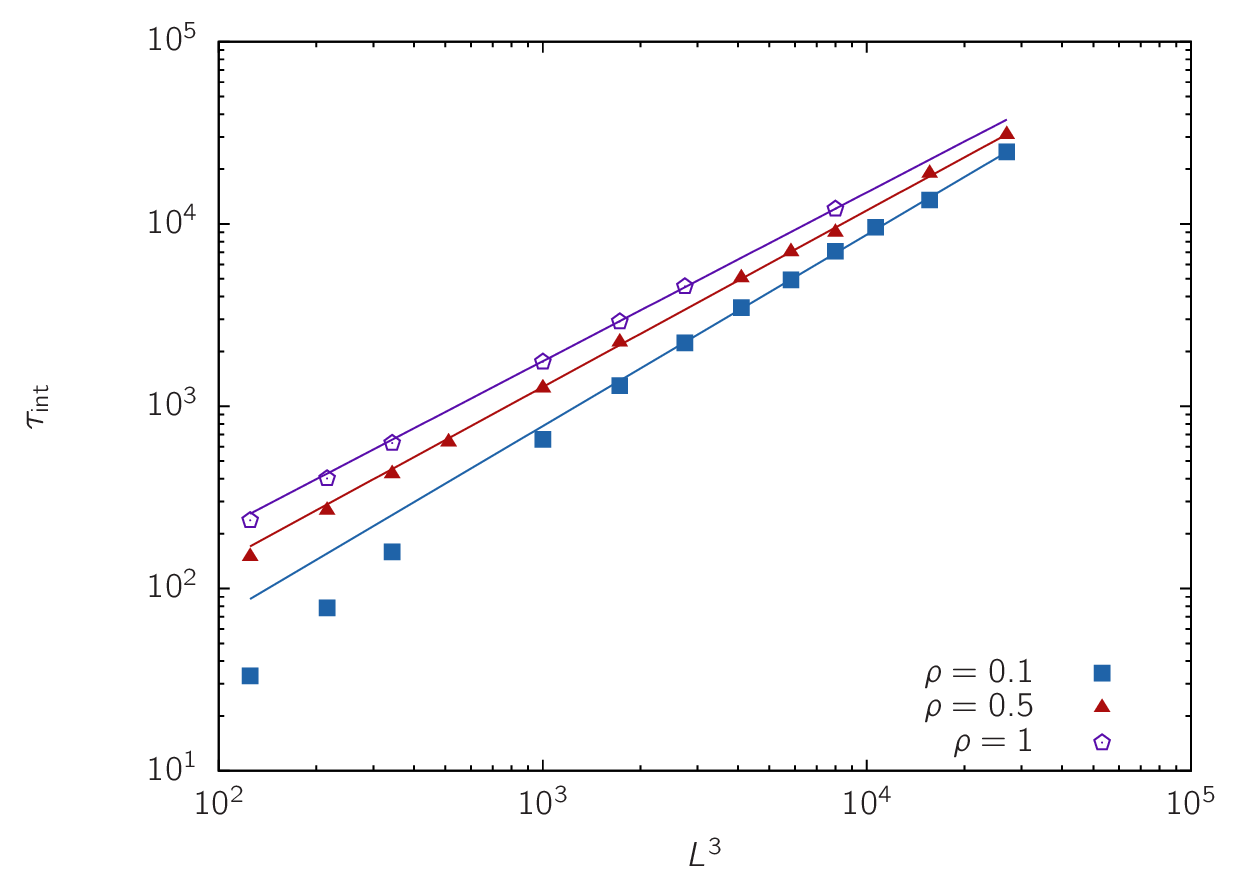}
 \caption{\label{fig:ac} Integrated autocorrelation times $\tau_{int}$ of compact polymers using the described algorithm. Integrated autocorrelation times show a power-law behavior yielding a linear dependence on $N$.} 
\end{figure}

As the algorithm only changes one bond per Monte Carlo step, subsequent conformations are highly correlated. A lot of subsequent Monte Carlo steps have to be performed until conformations get uncorrelated. Following Mansfield~\cite{Mansfield2006}, we calculate the autocorrelation function $C_{N_x}(t)$ of the observable $N_x$, which is defined as the number of bonds oriented along the $x$-direction. 
We then determine the exponential decay time of the correlations by a fit to the function $f(x) =\exp(-t/\tau_{exp})$ and obtain the exponential autocorrelation time $\tau_{exp}(\rho, N)$. This is the time  scale defining how long we have to wait initially before sampling any conformations.  On the other hand, the integrated autocorrelation time $\tau_{int}$ tells us how many Monte Carlo steps have to be carried out until we have a subsequent independent conformation. From Fig~\ref{fig:ac} one can see that $\tau_{int}(\rho, N)$ as a function of $N$ has a power-law behavior. We fit the correlation times to the function 
\[ \tau(\rho, N) = c(\rho) N^{d(\rho)} \]
and obtain the results shown in table~\ref{tab:ac}. Two conformations can be considered uncorrelated after $2\tau_{int}$ Monte Carlo steps~\cite{Binder2002}. Based on these results we write out a conformation after $14 N$ steps for $\rho=0.1$ and $6 N$ steps for $\rho > 0.1$. 

\begin{table}
\caption{\label{tab:ac} Fitting parameters for autocorrelation times. The exponential and integrated autocorrelation time was determined for different system sizes $L$. Then for each density used the obtained autocorrelation times were fitted to the function $\tau=cN^d$.}
\begin{tabular}{p{1cm}p{1.5cm}p{1.5cm}p{1.5cm}p{1.5cm}}\hline\hline
$\rho$ & $d (\tau_{exp})$ & $c (\tau_{exp})$ & $d (\tau_{int})$ & $c (\tau_{int})$\\\hline
$0.1$	& 1.07(1) & 5.07(36) & 1.05(1) & 6.2(5)\\
$0.5$	& 0.995(3) & 2.50(6) & 0.97(2) & 3.1(4) \\
$1$	& 0.954(3) & 2.32(5) & 0.926(4) & 2.94(10)\\\hline\hline
\end{tabular}
\end{table}

\section{Results}
In this section we present results on the conformational properties of compact polymers using the algorithm described above.
We performed extensive simulations for three different densities $\rho=0.1, 0.5$ and $1.0$. For each density a broad range of system sizes has been studied. For $\rho=0.1$ we did simulations up to $L=130$, for $\rho=0.5$ the largest system size studied is $L=80$ and for $\rho=1.0$ we were limited by computing time to  $L=55$. Thus the largest simulated chains are made up of $N=166\,375$ monomers for $\rho=0.1$, $N=256\,000$ for $\rho=0.5$ and $N=219\,700$ for $\rho=0.1$.  For each density and system size we sampled between $20\,000$ up to one million conformations, depending on the system size. 
\subsection{End-to-end distance statistics}
One characteristic length scale of a polymer is given by the mean squared end-to-end distance, i.e. the distance between the two endpoints of the chain, which is often denoted by $\left<R_e^2\right>$.  Obviously, $\left<R_e^2\right>$ depends on the total length of the chain, which is denoted by the number of monomers $N$ and in our case related to the system size $L$. The relation between the mean squared end-to-end distance and the number of monomers can be written in terms of a scaling law for polymers in good solvents,
\begin{equation}\label{eq:ete:dist:scaling}
 \left<R_e^2\right> \sim N^{2\nu}
\end{equation}
For random walks, i.e. polymers where excluded volume effects are ignored, it can be shown straightforwardly that the scaling exponent is $\nu=0.5$~\cite{Grosberg1994}. For self-avoiding walks in good solvents, where excluded volume effects are taken into account, the polymer is more swollen compared to the random walk. The resulting exponent is not known exactly, but estimated by field theoretical methods to $\nu\approx0.588$~\cite{Le1980}. A compact polymer on the other hand is characterized by an exponent of $\nu=1/3$ representing a globular shape with homogeneous density. The scaling law $\sqrt{\left<R_e^2\right>} = b N^{1/3}$ is also valid for the compact conformations studied here. Table~\ref{tab:ete:scaling} shows values for the parameter $b$ determined by a fit to the data for different densities $\rho$.

While a scaling law for the end-to-end distance distribution $P(r)$ for self-avoiding walks has been proposed long ago by Fisher~\cite{Fisher1966}, it remains unclear whether there is kind of universal scaling law for the end-to-end distribution  for compact polymers as well. Moreover it is not known what is the functional form of this scaling function. As $L$ is the only length scale in our system, which is related to $R=\sqrt{\left<R_e^2\right>} = b N^{1/3} = b \rho^{1/3}L$, there is a good chance that the distributions scale with $r/R$. This leads us to propose a scaling law for the distributions similar to that of a random walk or self-avoiding walk
\begin{equation} \label{eq:ete:pr_analytic}
 P(r) = \frac{A}{R} \left(\frac rR\right)^\mu \exp\left[-B \left(\frac{r}{R}\right)^\delta\right]
\end{equation}
For the random walk as well as the self-avoiding walk, the exponents $\mu$ and $\delta$ are well-known (see for example the book~\cite{Grosberg1994}). We now want to determine these exponents for the compact polymers studied here.

We can determine the parameters A and B by the following normalization conditions
\[ \int_0^\infty P(r) dr = 1 \qquad \int_0^{\infty} r^2 P(r) dr = R^2 \]
and obtain
\[ 
 B = \left[\frac{\Gamma\left(\frac{\mu+1}{\delta}\right)}{\Gamma\left(\frac{\mu+3}{\delta}\right)}\right]^{-\delta/2}\qquad
A = \delta\; \frac{B^\frac{1+\mu}{\delta}}{\Gamma\left(\frac{1+\mu}{\delta}\right)}
\]
Here $\Gamma(\cdot)$ denotes the Gamma function, which interpolates the factorial function. 

\begin{figure*}
 \includegraphics[width=\hsize]{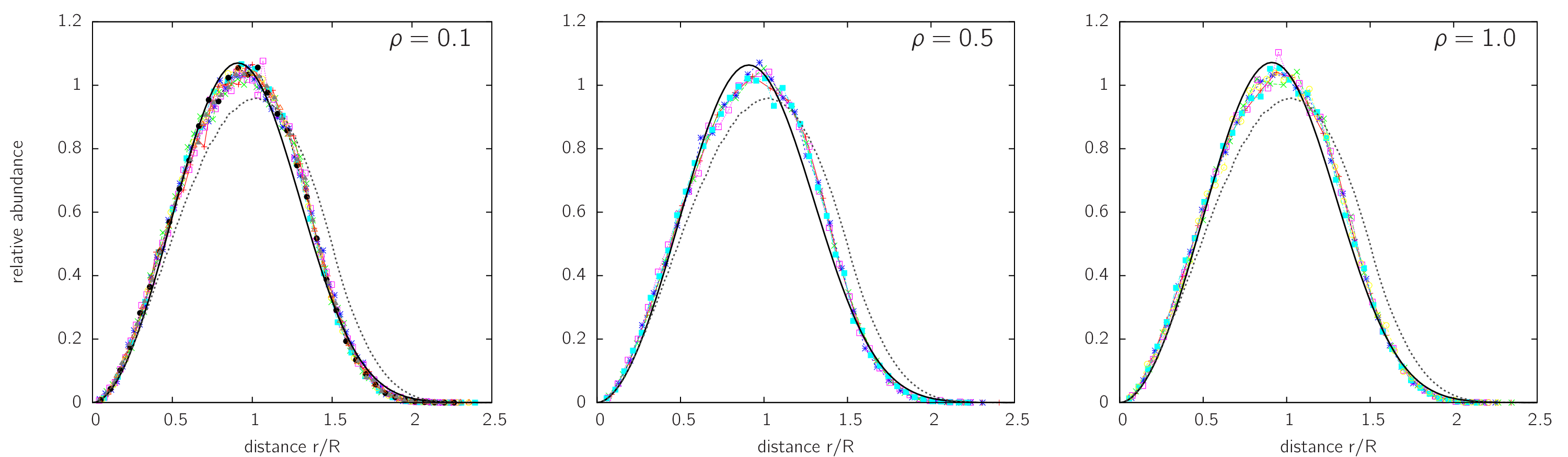}
 \caption{\label{fig:ete:pr} Probability distribution of the end-to-end distances for compact polymers. The distributions are scaled with root mean squared end-to-end distance $R=\sqrt{\left<R^2\right>}=aN^{1/3}$ where the parameter $a$ is determined by a fit according to table~\ref{tab:ete:scaling}. For each density systems of different size fall on top of each other. For $\rho=0.1$ system sizes vary from $L=10$ to $L=130$, for $\rho=0.5$ we analyzed systems from $L=10$ to $L=80$ and for $\rho=1$ the lattices used range from $L=10$ to $L=55$. The black lines represent a two-parameter fit to the empirical distribution function~\eqref{eq:ete:pr_analytic}, the fitting parameters are listed in~\eqref{eq:ete:prfitparams:all}. For comparison, the dotted line shows the distribution of two random points on a cubic lattice obeying excluded volume. }
\end{figure*}

In Fig.~\ref{fig:ete:pr} it is shown that the scaling with $R$ holds very well for different densities $\rho$. The analytic form of $P(r)$ approximates the data fairly well although deviations from the data are larger than for a self-avoiding walk, as the normalization condition above neglects the fact that end-to-end distances cannot extend beyond $\sqrt 2 L$. 

We fit the theoretical distribution function $P(r)$ to the data for three different $\rho$-values and obtain:
\begin{eqnarray}\label{eq:ete:prfitparams:all}
 \rho=0.1: \qquad \mu=1.89(3)\quad\delta=2.94(4) \\
 \rho=0.5: \qquad \mu=1.87(4)\quad\delta=2.91(6)\nonumber \\
 \rho=1.0: \qquad \mu=1.90(3)\quad\delta=2.94(5)\nonumber
\end{eqnarray}
Within the fitting error the exponents $\mu$ and $\delta$ are the same for different densities $\rho$, suggesting that these values show some universal features of compact polymers. On average we obtain from the above data
\begin{eqnarray}\label{eq:ete:prfitparams}
 \mu=1.889(65)\qquad\delta=2.932(89)
\end{eqnarray}

\begin{table}
\caption{\label{tab:ete:scaling} The effective monomer size determined by a fit to the function $\left<R_e^2\right>=b^2N^{2/3} = a^2L^2$ for each density $\rho$. Obviously, the effective monomer size decreases with increasing density.}
 \begin{tabular}{p{2cm}p{2cm}p{2cm}}
\hline\hline
  $\rho$ & $b$ & $a$ \\\hline
0.1 &    1.50(2)	&	0.696(7)\\
0.5  &   0.89(7)	&	0.712(5)\\
1    &   0.71(1)	&	0.71(1)\\\hline\hline
 \end{tabular}
\end{table}

\begin{figure}
 \includegraphics[width=\hsize]{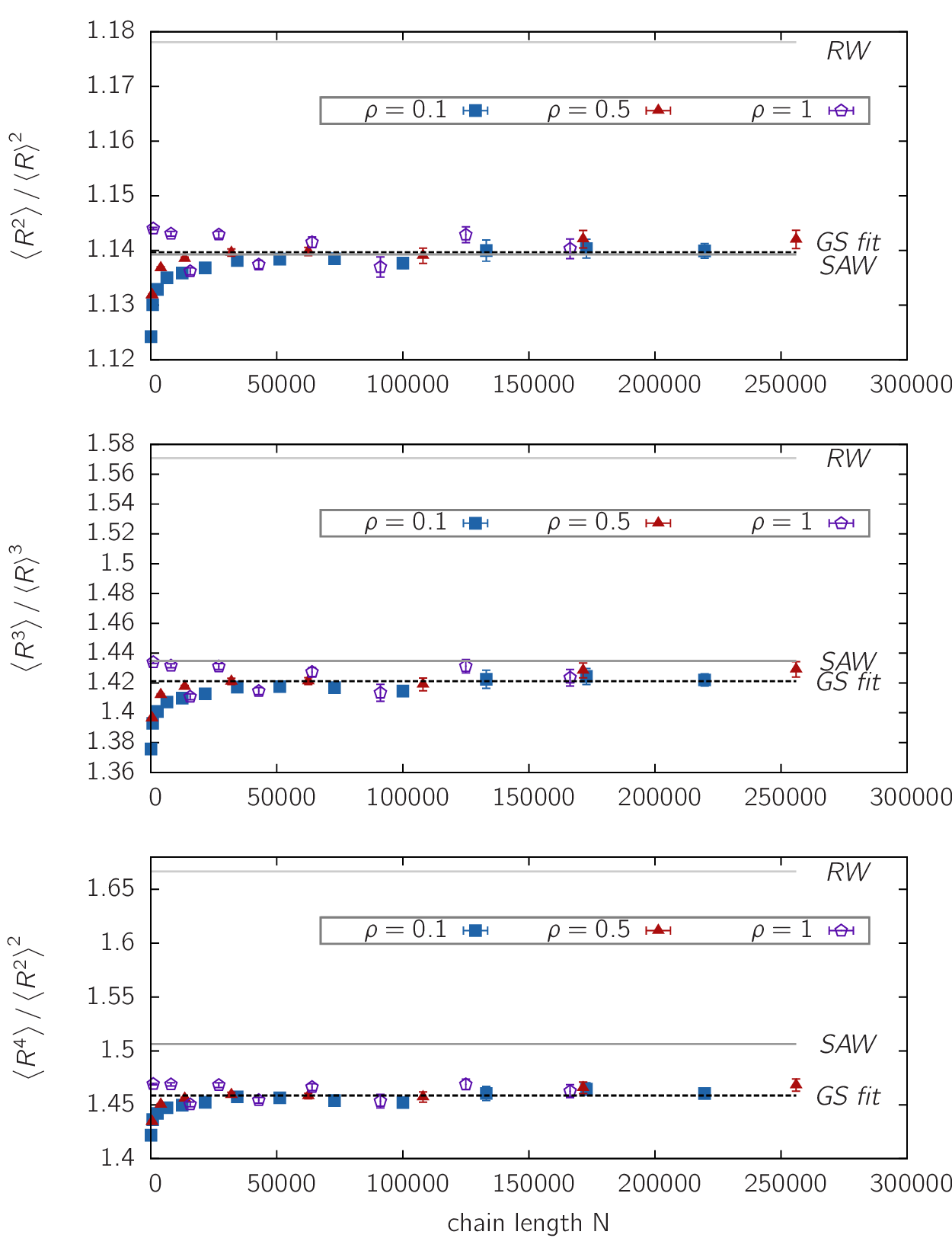}
  \caption{\label{fig:ete:moments} Ratios of the moments of the end-to-end distances. The ratios are independent of the system size $L$ (except finite size effects) and seem to be independent of the density $\rho$.}
\end{figure}

Comparing experimental data to the distribution function $P(r)$ is not always the method of choice, especially when the number of data points is too small for creating reasonable histograms. In this case it is more useful to look at the first few moments of the distribution. Here we analyze dimensionless ratios of moments of the end-to-end distance distribution, having the advantage that no adjustable parameter is present. The ratios of interest here are
\[ c_2 = \frac{\left<R^2\right>}{\left<R\right>^2}, \qquad c_3=\frac{\left<R^3\right>}{\left<R\right>^3}, \qquad c_4=\frac{\left<R^4\right>}{\left<R^2\right>^2} \]
For a random walk and a self-avoiding walk, the $c_i$ are constants not depending on any model parameters (such as linker length $l$). Here we show that this is also the case for compact polymers and we determine its values. The ratio plots are shown in Fig.~\ref{fig:ete:moments}. A fit yields the values
\[ c_2 = 1.1395(5) \qquad c_3=1.421(2) \qquad c_4= 1.458(1)\]
Obviously, there are significant differences from a RW, however differences from a SAW behavior become only visible in the fourth order ratio $c_4$. We demonstrate in a later section how this information can be used to characterize the behavior of biopolymers. 

Instead of determining the moments by the raw simulational data we can determine the moments by the analytic function $P(r)$ given in eq.~\eqref{eq:ete:pr_analytic}. Calculations yield
\begin{eqnarray}
 c_2 &=& \frac{\Gamma\left(\frac{3+\mu}{\delta}\right)\Gamma\left(\frac{1+\mu}{\delta}\right)}{\Gamma\left(\frac{2+\mu}{\delta}\right)^2} \\
 c_3 &=& \frac{\Gamma  \left( {\frac {{\it \mu}+4}{{\it \delta}}} \right)  
\Gamma  \left( {\frac {{\it \mu}+1}{{\it \delta}}} \right)  ^{2}}{
  \Gamma  \left( {\frac {2+{\it \mu}}{{\it \delta}}} 
 \right) ^{3}} \\
 c_4 &=& \frac{\Gamma  \left( {\frac {{\it \mu}+1}{{\it \delta}}} \right) \Gamma 
 \left( {\frac {{\it \mu}+5}{{\it \delta}}} \right)}{ \Gamma 
 \left( {\frac {3+{\it \mu}}{{\it \delta}}}  \right)^{2}}
\end{eqnarray}
Plugging in the fit parameters for $\mu$ and $\delta$ from eq.~\eqref{eq:ete:prfitparams} we obtain
\[ c_2=1.140(6) \qquad c_3=1.43(2)\qquad c_4=1.488(24)\]
The values are compatible with the ones calculated directly by a fit to the simulation data within the range of the errors. Deviations become large for larger moments reflecting the approximative character of the scaling function $P(r)$. 

\subsection{Intrachain distance statistics}
In the past a lot of effort has been undertaken to study the end-to-end distributions of self-avoiding walks~\cite{Fisher1966}. For random walks, this problem is easily solved analytically. In the last section we studied these distributions for compact polymers which has not been done so far. We found a similar scaling function as for random and self-avoiding walks with exponents which seem to be universal for compact polymers of different densities and chain lengths. However, from the experimental point of view one is not only interested in the distance between end points of a compact polymer but also in the distance distribution between two arbitrary monomers along the chain which are separated by a certain contour length $n$. For example, this quantity becomes important in experiments measuring the spatial arrangement of two flourescently labelled parts of the human genome~\cite{Mateos2009}. Therefore, we have to evaluate how the distribution changes when looking at intrachain segments. 

\begin{figure}
 \includegraphics[width=\hsize]{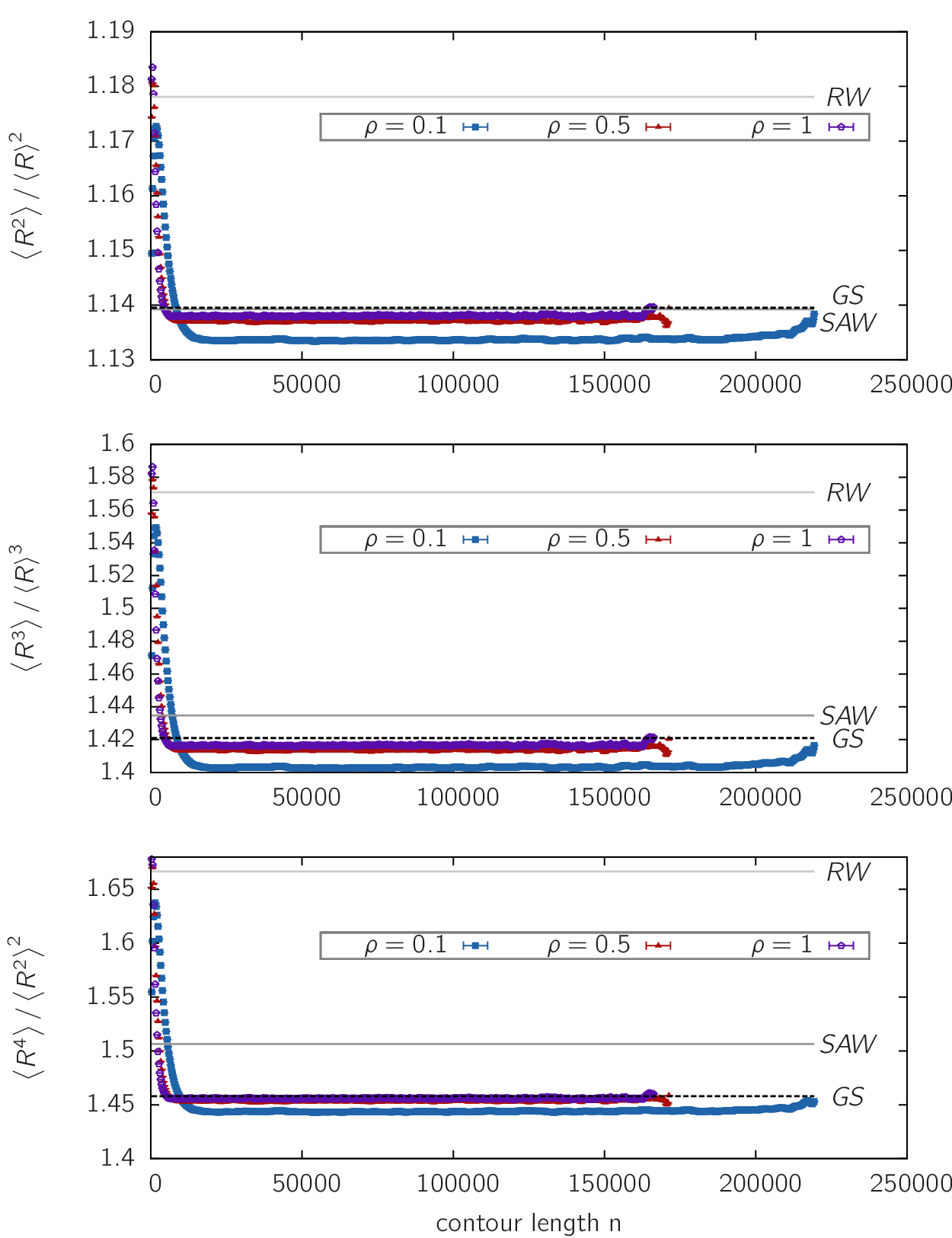}
 \caption{\label{fig:ic:moments} This figure shows the ratios of the moments of the distribution of distances between two parts of the chain which are separated by a contour length $n$.}
\end{figure}

It is most interesting to look at the moment ratios as these ratios are easiest to compare to experimental data, which quite often do not provide enough data points to obtain a complete distribution function. For the largest chains simulated the moment ratios are shown in Fig.~\ref{fig:ic:moments} for various contour length $n$. The mean values are averages both over different positions along one chain as well as over the set of sampled conformations $\mathcal{C}$. The $k$th moment is thus evaluated as
\begin{equation}
\left<R_n^k\right>  = \frac{1}{|\mathcal{C}|} \frac{1}{N-n}\sum_{C\in\mathcal{C}}  \sum_{i=1}^{N-n} \parallel \mathbf{r}_{i+n}^C - \mathbf{r}_{i}^C\parallel^k 
\end{equation}
Here $\mathbf{r}_i^C$ denotes the position of the $i$th monomer of the conformation $C$ out of the set of sampled conformations $\mathcal{C}$. 

For small contour length $n$ the moment ratios are peaked and reach the value of a random-walk. This is due to the screening effect in compact polymers, which is in detail analyzed in one of the following sections. However, the ratios pretty fast fall below the ratios for self-avoiding walks and stay mostly constant. There is only a small increase for the ratios where the contour length approaches the chain length $N$, indicating that the chain ends have more freedom for fluctuations than parts embedded in the middle of the chain. We want to stress here that for sufficiently large contour lengths the moment ratios for intrachain distances are smaller than those for the end-to-end distances.

\subsection{End-point statistics}
It is an open question, raised for example in Ref.~\cite{Lua2004}, whether the positions of the end points of a compact polymer are correlated or not. It was suspected that an entropic cost associated with local rearrangement around the chain's ends might cause some effective attraction or repulsion between them. Let $\mathbf{r}_1=(x_1, y_1, z_1)$ and $\mathbf{r}_2=(x_2, y_2, z_2)$ denote the end point vectors with respect to the center of the simulation box. Lua et al.~\cite{Lua2004} showed then for Hamiltonian paths up to $L=10$ that the end point correlation coefficient
\begin{equation}
\label{eq:endpoint:correlation} c=\frac{\left<x_1x_2\right>}{\sqrt{\left<x_1^2\right>\left<x_2^2\right>}}
\end{equation}	
is negative for small lattice sizes but pretty fast approaches the correlation between disconnected points only obeying excluded volume and the chess board theorem~\cite{Lua2004}. The latter theorem states that, if we mark adjacent vertices on the lattice graph with different colors similar to a chess board, then the end points of a chain with even numbers of monomers are sitting on lattice sites with different color while the end points of a chain with an odd number of monomers are positioned on lattice sites of same color. This restriction has to be taken into account when comparing to randomly positioned points, as this is an inherent feature of the lattice model but not of the ensemble of compact polymers in general. However it becomes more and more negligible the larger the lattice size.

\begin{figure}
 \includegraphics[width=\hsize]{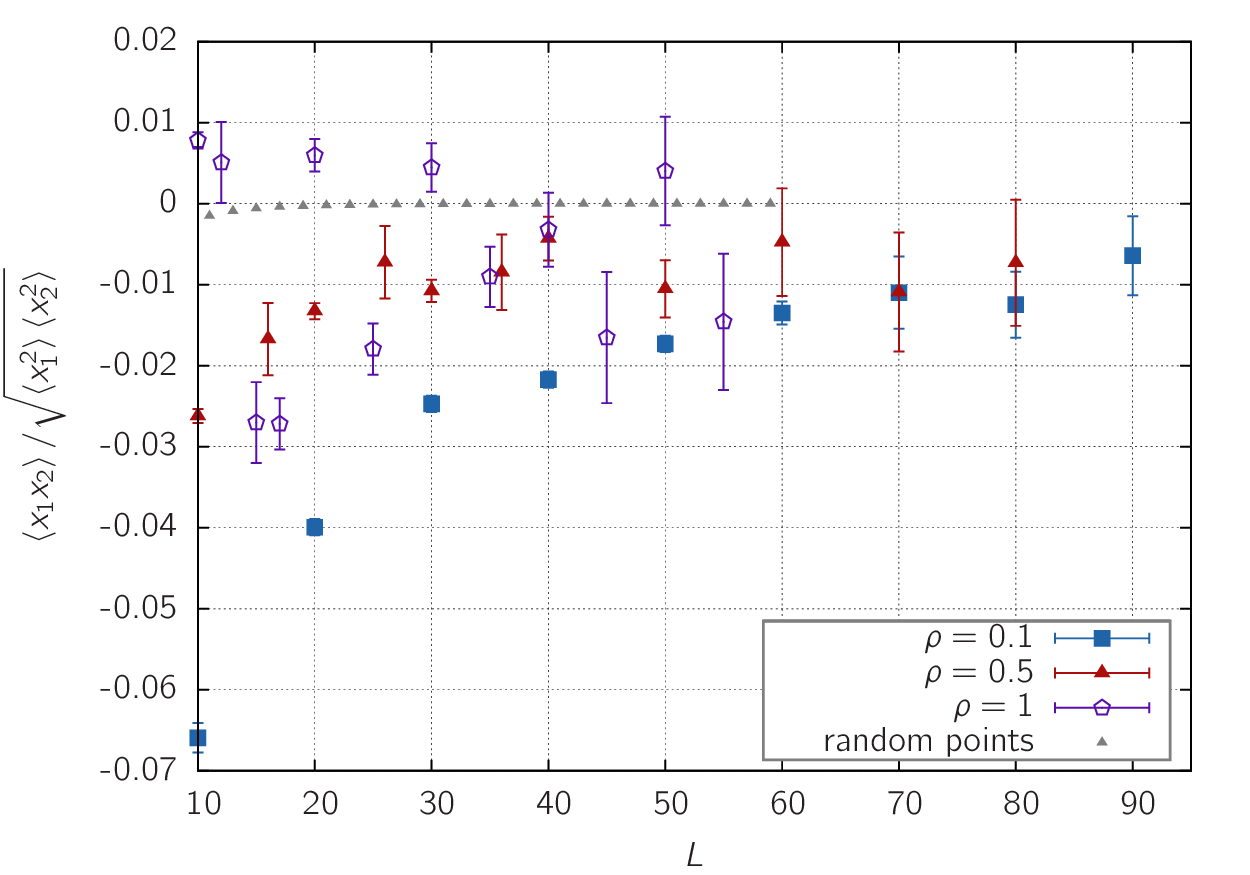}
 \caption{\label{fig:endpoints:corr} End-end correlation coefficients according to eq.~\eqref{eq:endpoint:correlation} for compact polymers with different densities ($\rho=0.1$, $\rho=0.5$ and $\rho=1$). Additionally shown are exact results for the correlation coefficient of two randomly positioned points on a cubic lattice only obeying excluded volume and the chess board theorem.}
\end{figure}

Here we study the correlation coefficient $c$ for lattice sizes much larger than in~\cite{Lua2004}. Note that the coordinates in eq.~\eqref{eq:endpoint:correlation} are taken with respect to the center of the simulation cube. Fig.~\ref{fig:endpoints:corr} shows that there are negative correlations for all densities considered, which approach zero for larger lattice sizes. The only deviations are for $\rho=1$ (i.e. Hamiltonian paths) and even lattice sizes, which obviously is an effect of the lattice geometry and the chess board theorem and therefore no intrinsic property of compact polymers. These results are in very good agreement with the results by Lua et al.~\cite{Lua2004}, suggesting that there are no end-point correlations in the $N\rightarrow\infty$ limit. 

\begin{figure}
 \includegraphics[width=\hsize]{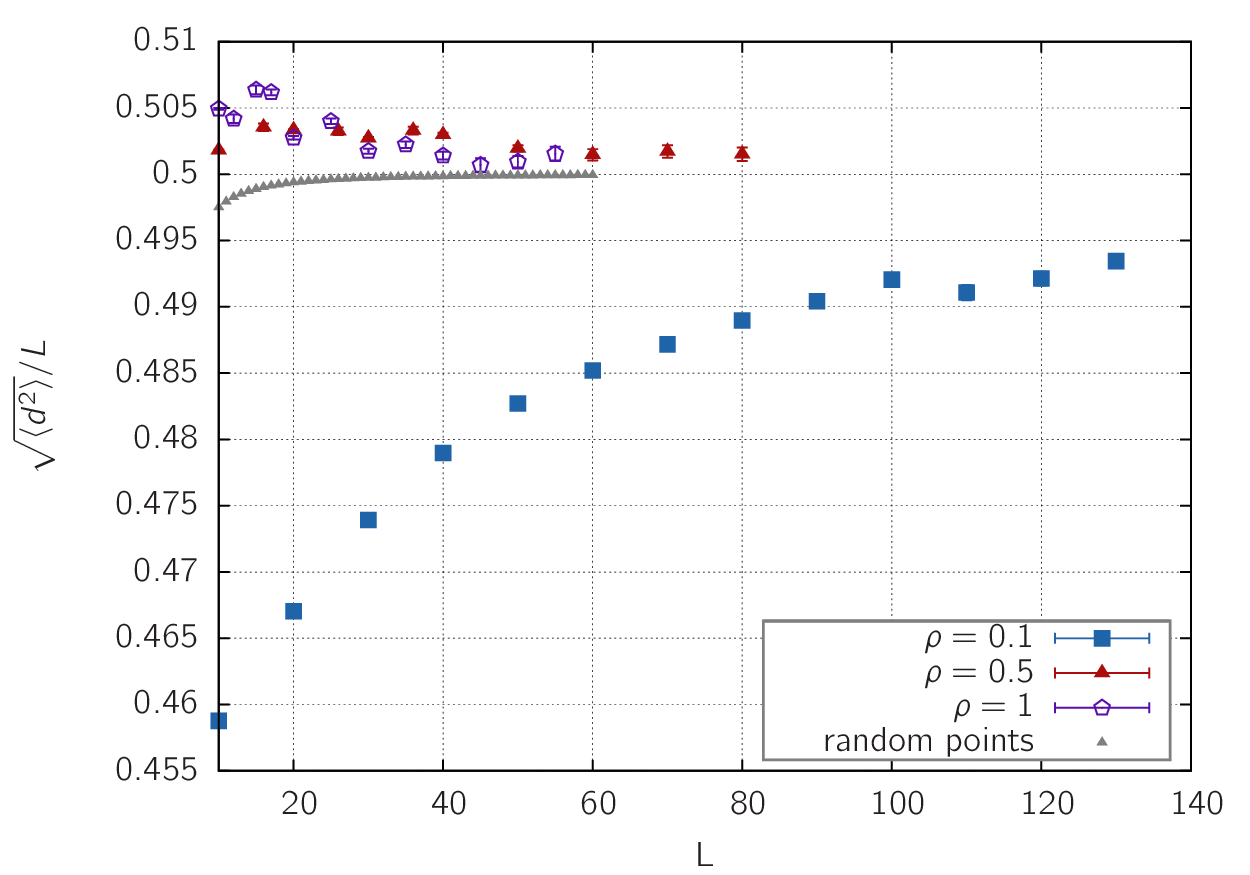}
 \caption{\label{fig:endpoints:msd} Distance of the end points from the center of the cube for compact polymers with different densities. For comparison shown are also the results for random points obeying excluded volume and the chessboard theorem.}
\end{figure}

We also analyzed the mean square displacement of the endpoints from the center of the cube in order to answer the question whether the polymer tends to arrange such that the endpoints predominantly locate in the center of the cube or at its periphery. Being located in the center of the cube might be disadvantageous due to entropic reasons. Fig.~\ref{fig:endpoints:msd} shows that for $\rho=0.5$ and $\rho=1$ the points are predominantly shifted towards the periphery of the cube, while for $\rho=0.1$ the points are located more in the interior of the cube. ~

\subsection{Correlations of intrachain segments}
\begin{figure}
\includegraphics[width=\hsize]{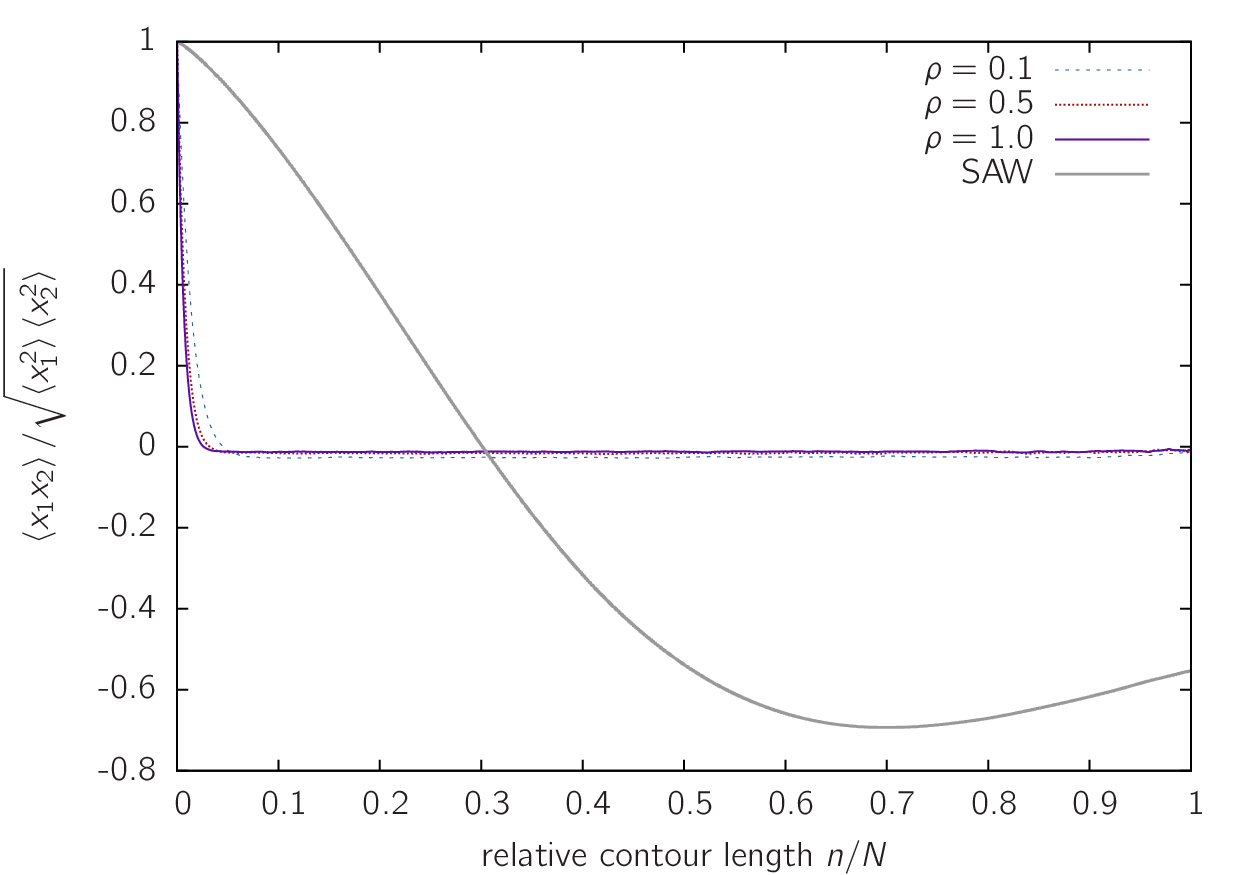}
\caption{\label{fig:correlations} Position correlations of monomers. This figure shows the correlation coefficient $\frac{\left<x_1x_2\right>}{\sqrt{\left<x_1^2\right>\left<x_2^2\right>}}$ between two monomers in dependence of their separation along the chain. The data shown are for system sizes $L=55 (\rho=1), L=70 (\rho=0.5)$ and $L=130 (\rho=0.1)$.For comparison the correlations in a self-avoiding walk chain of length $N=10\,000$ are also shown.}
\end{figure}

Consider two arbitrary points on a polymer with coordinates $\mathbf{r}_1 = (x_1, y_1, z_1)$ and $\mathbf{r}_2 = (x_2, y_2, z_2)$. As for the end points of the chain, we can pose the question whether the coordinates of these points are correlated by evaluating the correlation coefficient of eq.~\eqref{eq:endpoint:correlation}
We assume the coordinates $\mathbf{r}_1$ and $\mathbf{r}_2$ to be given with respect to the center of mass of the polymer. A value of $c=0$ indicates that there is no correlation between the coordinates $\mathbf{r}_1$ and $\mathbf{r}_2$, i.e. they effectively behave like two randomly chosen points on the cubic lattice. A value $c\neq0$ indicates an effective attraction or repulsion.  Fig.~\ref{fig:correlations} shows the correlation coefficient in dependence of the contour length $n$ between the segments. While for short contour length there are high correlations because of the connectivity of the chain, these correlations decay fast and for larger contour length correlations are nearly vanished. This is in stark contrast to the behavior of a self-avoiding walk, where (negative) correlation effects are dominant even on the length of the whole chain. This result is in perfect agreement with the scaling theory developed by De Gennes~\cite{Gennes1979}, which predicts that on the length scale of the compact system, parts of the chain become practically independent. If the scaling theory is correct then the decay length of the correlations $n_d$, which we define as the length where $c(n_d)\sim 1/e$,  should be related to the system size $L$. We test this prediction by evaluating the ratio $r = \left<R^2(n_d)\right>/L^2$ where $\left<R^2(n_d)\right>$ is the mean square displacement between the end points of a segment of length $n_d$.  For the largest system sizes studied we find the values $r\approx0.31$ for $\rho=1$, $r\approx0.31$ for $\rho=0.5$, $r\approx0.29$ for $\rho=0.1$. As $r$-values are nearly equal for systems of completely different size and density there is strong evidence that the decay of the position correlations is directly related to the system size.

\subsection{Screening of excluded volume in compact polymers}
Much is known about polymer melts, where a number of polymers with degree of polymerization $N$ is placed in a system with volume fraction $\Phi$. Below a critical concentration (or volume fraction) $\Phi^\star$, the polymers do not feel the existence of the others and basically behave like self-avoiding walks in a good solvent. This critical concentration $\Phi^\star$ is given by the volume fraction where the free coils with extension $\left<R^2\right>_0\sim N^{2\nu}$ begin to overlap~\cite{Gennes1979}. The index $0$ indicates here the limit $\Phi \rightarrow 0$. The value of the critical concentration scales like 
\[ \Phi^\star \approx \frac{N}{ \left<R^2\right>_0} \sim N^{-(3\nu-1)} \]
 At volume fractions above $\Phi^\star$ the polymers begin to feel each other and the system can be described  by the correlation length $\xi$. On scales larger than this correlation length, the chains effectively behave like ideal coils, a theorem most often referred to as Flory theorem. On a scale smaller than $\xi$ excluded volume effects still play a dominant  role. By scaling arguments one finds~\cite{Gennes1979} that
\[ \xi ~\sim \Phi^{-\nu/(3\nu-1)} \]

Here we want to investigate the question whether there is a fundamental difference between a polymer melt with volume fraction $\Phi$ and a compact polymer, i.e. a melt with degree of polymerization $N=\Phi L^3$. Consider an arbitrary segment of a compact polymer of length $N_m$. We impose on $N_m$ the condition that the extent of this segment must be smaller than the system size in order to avoid effects of the confinement to play a role. We now ask whether these segments of length $N_m$ of our compact polymers behave different from a polymer with degree of polymerization $N_m$ in a corresponding melt. The analysis of positional and angular correlation effects which are decaying pretty fast suggests that a part of the chain should not ``feel'' that it is connected with a part far away.

\begin{figure}
 \includegraphics[width=\hsize]{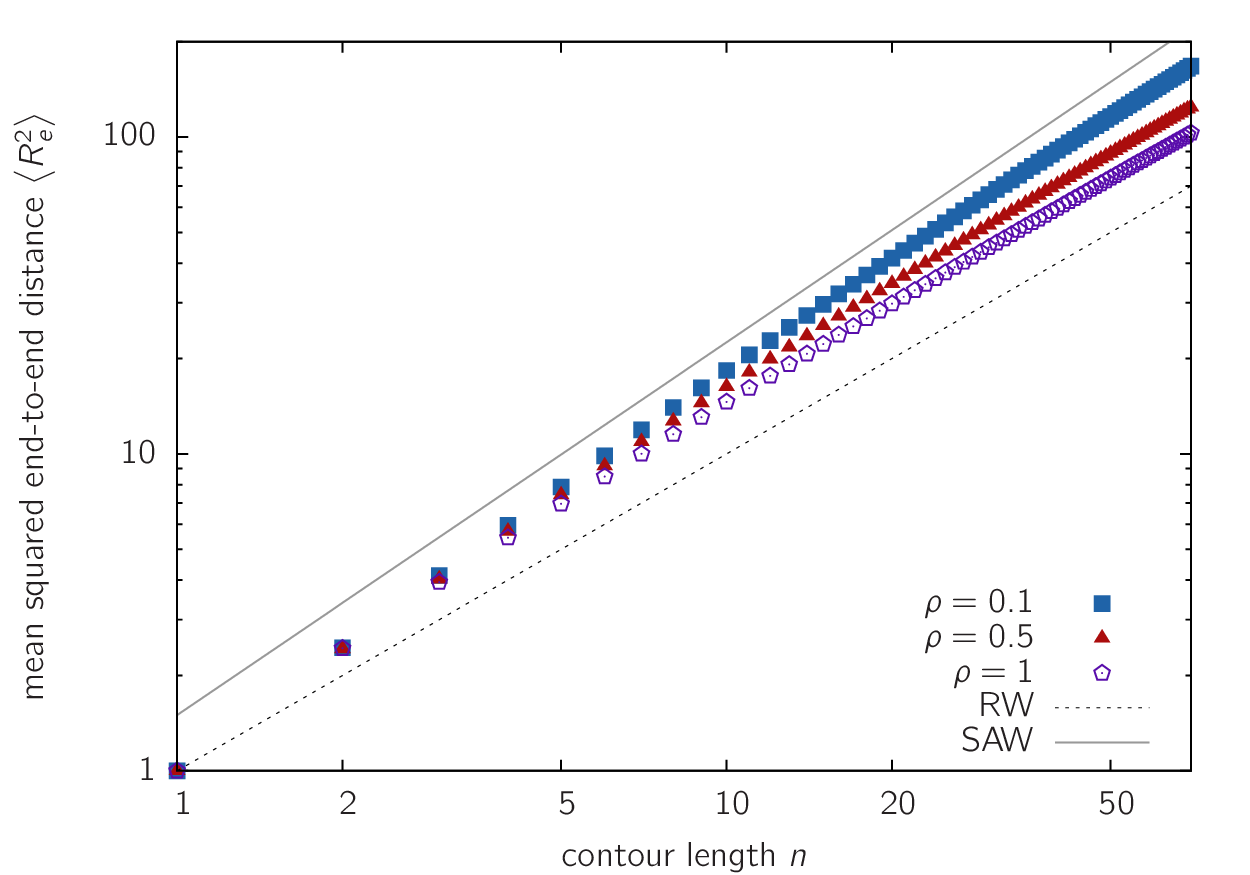}
 \caption{\label{fig:screening:r2} Mean squared distances $\left<R_n^2\right>$ between monomers separated by a contour length of $n$ for different densities. The compact polymers also show a screening of excluded volume for large densities.}
\end{figure}

The theory of polymer solutions predicts that there is a crossover from a self-avoiding walk behavior to a random walk behavior on the short scale~\cite{Paul1991}. Fig.~\ref{fig:screening:r2} shows that this crossover becomes indeed apparent in the mean square displacement for short contour lengths in compact polymers. While for $\rho=0.1$ we find a self-avoiding walk type of scaling with exponent $\nu=0.588$ for contour lengths up to $n\approx 50$, the maximal dense system behaves more like a random walk with $\nu=0.5$ indicating the screening of excluded volume in this system. 

To analyze the screening length we have to look in more detail at the structure function of parts of the chains. The structure function is defined as 
\begin{equation}
 \label{eq:screening:sq}
 S(q) = \left< \frac{1}{N_m} \left| \sum_{j=0}^{N_m} e^{i\mathbf{q}\cdot\mathbf{r}_j}\right|^2\right>_{q}
\end{equation}
The brackets denote a spherical average over all $\mathbf{q}$-vectors of equal magnitude and over all conformations. The sum is over a subchain of length $N_m$ whose position vectors are denoted by $\mathbf{r}_0\ldots\mathbf{r}_{N_m}$. One expects these subchains to behave like random walks on distances larger than $\xi$ and self-avoiding walks on distances smaller than $\xi$, i.e.
\begin{equation}\label{eq:screening:sqscaling}
 S(q) \sim \begin{cases} q^{-2}, \qquad \left<R^2\right>^{1/2} > \frac{2\pi}{q} > \xi \\
              		   q^{-1/\nu}, \qquad \xi  > \frac{2\pi}{q} > 1
             \end{cases}
\end{equation}

\begin{figure}
\ifthenelse{\lengthtest{\hsize>12cm}}{\includegraphics[width=0.5\hsize]{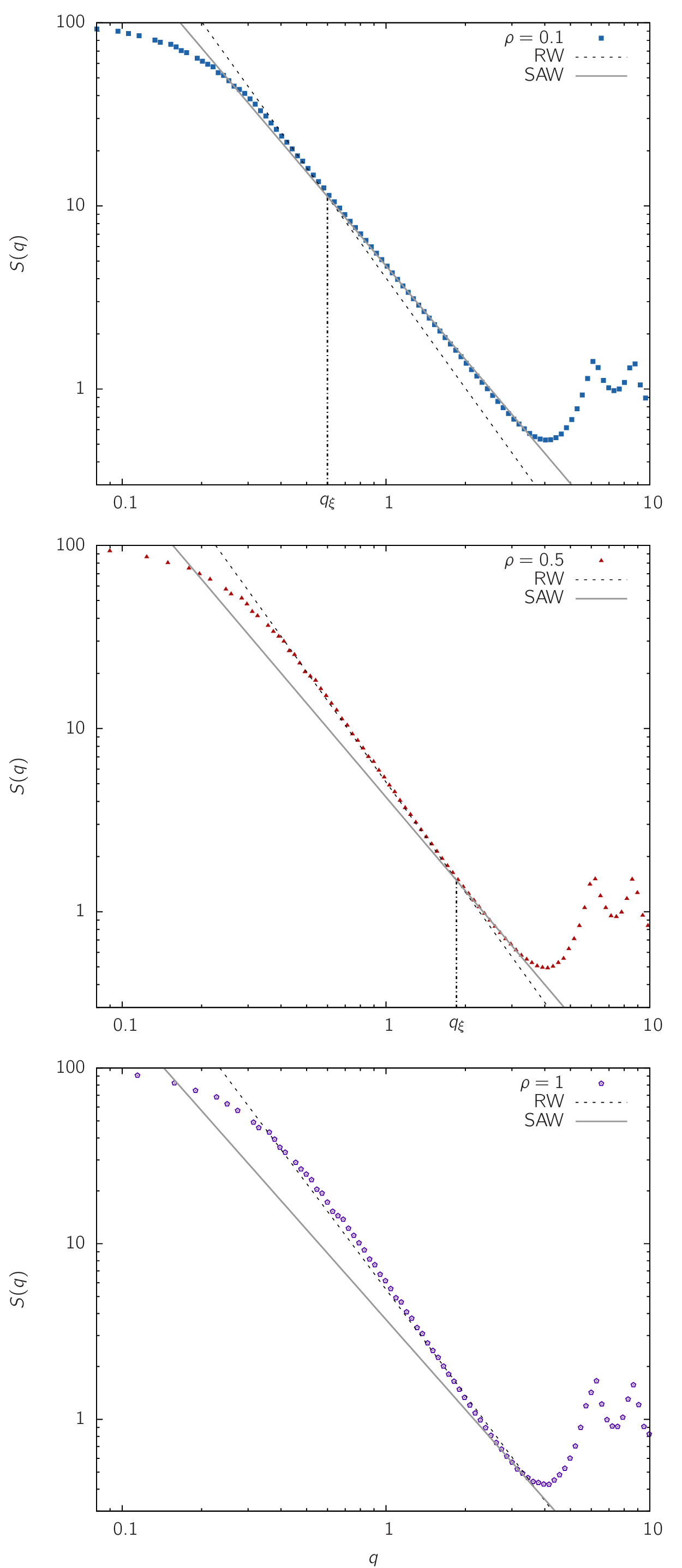}}{\includegraphics[width=\hsize]{sq_n100_combined_long.ps}}
 \caption{\label{fig:screening:sq} Structure function of chain segments of length $N_m=100$ for compact polymers with different densities. For the calculation the largest simulated system sizes for each density was used. The value of the segment length was chosen such that the radius of gyration of the segments is way below the system size. Shown is also the scaling regimes where $S(q)\sim q^{-2}$ and $S(q)\sim q^{-1/v}$. From a fit to the curves one can determine the crossover value $q_{\xi}$ which determines the screening length.}
\end{figure}

In Fig.~\ref{fig:screening:sq} the structure function is shown for chain segments of length $N_m = 100$. One can immediately see that the system with $\rho=0.1$ shows a range of $q$-values, where excluded volume is not screened, extending much beyond the length scale of a single bond. On the other hand for $\rho=0.5$ excluded volume interactions are screened very fast resulting in ideal chain behavior over a wider range of $q$-values. We can determine the screening length by performing a linear fit with slope $-2$ for small $q$-values (but beyond the scale where asymptotic behavior sets in) and  a linear fit with slope $-1/\nu$ for large $q$-values (but away from the length scale of a single bond).  We then extract the value $q_{\xi}$ where the crossover between the two regimes occurs and obtain the following screening lengths
\begin{align*}
 \rho=0.1 &:&\quad q_{\xi}\approx 0.6 &\quad\rightarrow\quad& \xi=\frac{2\pi}{q_{\xi}} \approx&\: 10.5 \\
 \rho=0.5 &:&\quad q_{\xi}\approx 1.8 &\quad\rightarrow\quad& \xi=\frac{2\pi}{q_{\xi}} \approx&\: 3.5 \\
 \rho=1.0 &:&\quad  &\quad\quad& \xi=\frac{2\pi}{q_{\xi}} \approx&\: 1 
\end{align*}
For $\rho=1$ the system seems to be that dense that excluded volume is shielded on the order of a bond length, therefore the scaling regime where $S(q)\sim q^{-1/\nu}$ does not show off any more. This result does not come as a big surprise as the screening length $\xi$ is related to the average mesh size in the system~\cite{Gennes1979}. This mesh size is -- as every lattice site is occupied -- approximately equal to unity. 

\subsection{The gyration tensor}
\begin{figure}
 \includegraphics[width=\hsize]{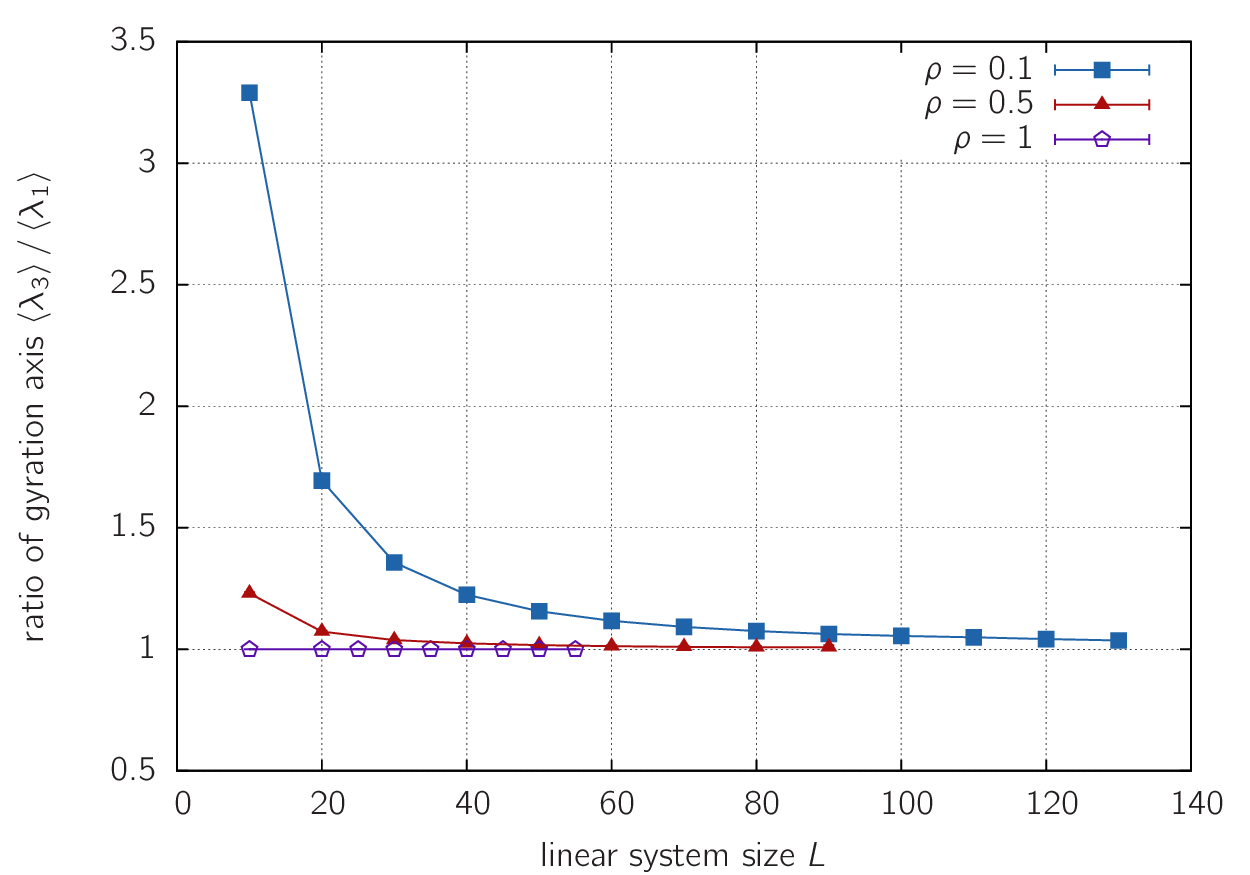}
 \caption{\label{fig:gyr:ratios} The ratio between the largest and the smallest eigenvalue of the gyration tensor. Eigenvalues are the squared lengths of the gyration ellipsoids main axes. While for RW or SAW polymers the tensors show a characteristic asphericity, compact polymers have an almost spherical shape in the asymptotic limit.}
\end{figure}

The shape of a a polymer is described by its gyration tensor. The gyration tensor is defined as 
\begin{equation}
 S_{mn} = \frac{1}{N} \sum_{i=1}^N r_m^{(i)} r_n^{(i)}
\end{equation}
Here $\mathbf{r}^{(i)}$ is the coordinate vector of the $i$th monomer and the subindex denotes its cartesian components. The eigenvalues $\lambda_1 \le \lambda_2 \le \lambda_3$ give the squared lengths of the principal axes of gyration. The ratios of the eigenvalues indicate the deviation  from a sphere-like shape of the polymer. 
It is well-known, for example, that the gyration tensor for self-avoiding walks and random walks in good solvent has a pronounced asphericity. This asphericity shows up in the asymptotic ratios of the eigenvalues, namely~\cite{Bruns1992, Solc1971}
\begin{align*}
 \left<\lambda_3\right> : \left<\lambda_2\right>: \left<\lambda_1\right> &\rightarrow 12 : 2.7 : 1  &\quad& &\text{for a RW}\\
 \left<\lambda_3\right> : \left<\lambda_2\right>: \left<\lambda_1\right> &\rightarrow 14 : 2.98 : 1 &\quad& &\text{for a SAW}
\end{align*}
For Hamiltonian paths it is clear that there can be no asphericity in a symmetric simulation box as every lattice site is occupied. However it is not clear \textit{a priori} that this holds true for less dense systems, where a crossover to the self-avoiding walk behavior might occur. 
Fig.~\ref{fig:gyr:ratios} shows the ratio $\left<\lambda_3\right>:\left<\lambda_1\right>$. For $\rho=0.1$ a pronounced deviation from the symmetry shows up for small $N$, but evidently, this deviation vanishes for large system sizes. 

\section{Application of the results to experimental data}
In this section we apply the findings from the previous sections to a current research field of biophysics: the folding of chromatin. Chromatin is a compacted state of DNA, which can be found inside the nucleus of eukaryotic cells~\cite{Schiessel2001}. This compacted state arises when the DNA double strand is wrapped around histone cores forming the so-called nucleosomes which in turn arrange in a beads-on-a-string-like manner forming the chromatin fiber.  Amazingly little is known about the higher-order organization of chromatin inside the interphase nucleus. Indeed, about 2 meters of double-stranded DNA have to be densly packed into a nucleus of only 10 $\mu m$ in diameter while still parts of the DNA have to be accessible for large proteins. The detailed folding mechanisms are still a mystery. One reason for the slow advance in the field of chromatin folding is that imaging techniques do not allow one to follow the chromatin fiber along its contour inside the living cell. High-resolution techniques like EM are too invasive to maintain the structure one wants to observe. Therefore, one has to rely on indirect approaches. One such approach is fluorescence in situ hybridization (FISH). Here two fluorescent markers are positioned along the contour of the chromatin fiber having a certain genomic distance $g$ between them. Confocal light microscopy then allows one to determine the spatial distance between these two markers. 

In an earlier study~\cite{Bohn2007a} we presented a model for chromatin organization, the random loop (RL) model. This model assumes that random chromatin-chromatin interactions play a dominant role for the structural organization of chromatin in the interphase nucleus. The random loop model predicts a leveling-off in the mean square displacement between two FISH markers in relation to the genomic distance. This has been nicely confirmed by experiments~\cite{Mateos2009}. Importantly, the RL model explains the folding of chromatin without assuming a confined geometry. This is consistent with experimental data showing that the leveling-off takes place well below the diameter of the cell nucleus. 

Here we want to reanalyze the experimental data keeping an eye on the type of compact polymers studied above. Obviously, compact polymers also show a leveling-off in the mean square displacement. With respect to Hamiltonian paths this is necessary, as there is the confined space of the cubic box. However this is a general feature of compact polymers. Therefore the question may be asked whether there is a difference between the random loop polymers and generic compact polymers. Indeed, compact polymers arise whenever attractive interactions begin to play a dominant role in comparison to entropic forces. The random loop model also assuming attractive intra-chromatin interactions. The important difference is that the random loop ensemble contains an average over a disorder given by the randomness in the loop attachment points. 

\begin{figure}
  \ifthenelse{\lengthtest{\hsize>12cm}}{\includegraphics[width=0.5\hsize]{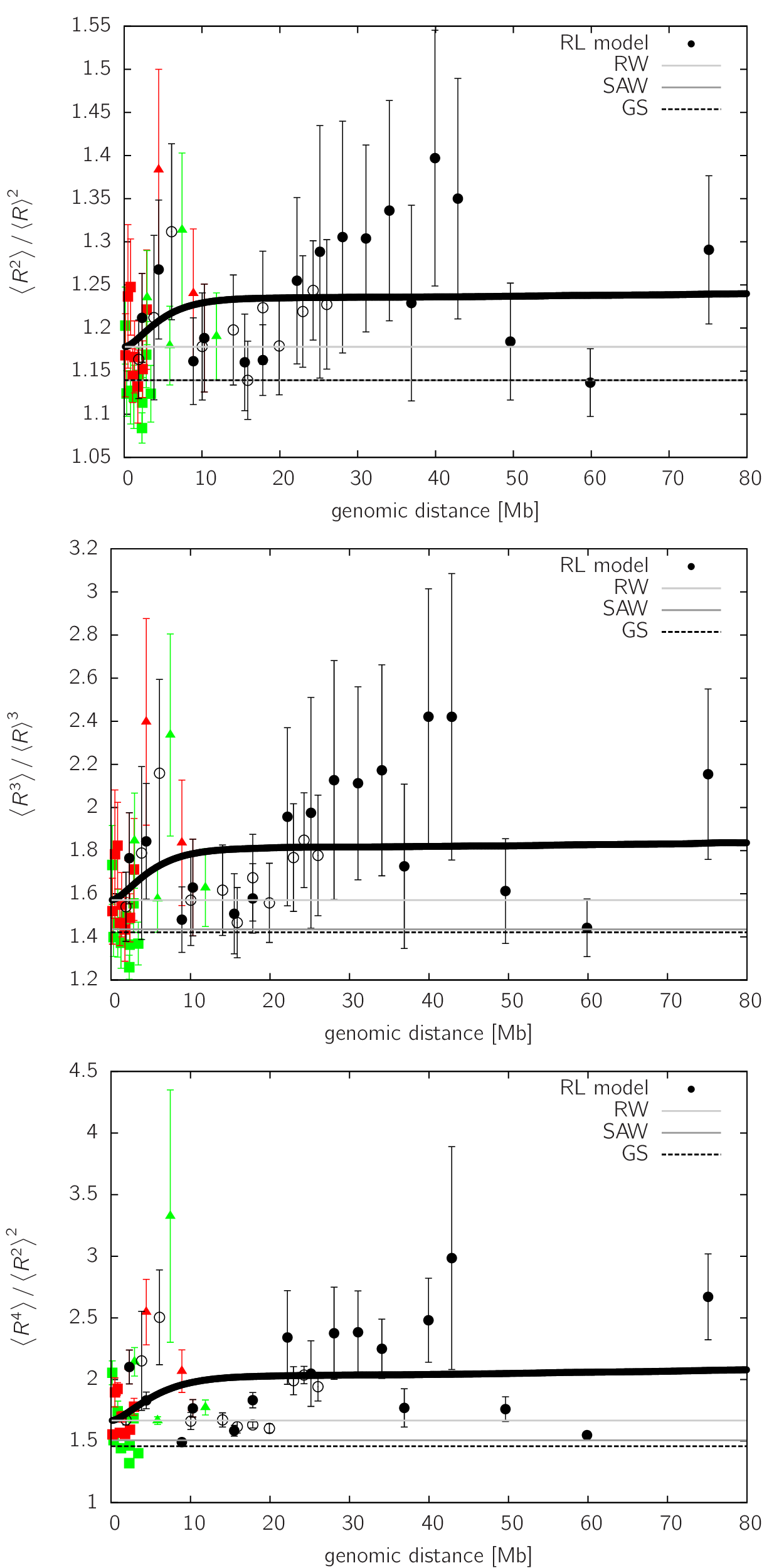}}{\includegraphics[width=\hsize]{experimental_moments.ps}}
 \caption{\label{fig:exp:moments} Experimental data from FISH measurements~\cite{Mateos2009}. Shown are the moment ratios of the measurements together with the predicted values for the random walk (RW), the self-avoiding walk (SAW) and the globular state (GS). Obviously the data has significantly higher moment ratios than these classical polymer models. The RL model comes closer to ratios of the experimental data by virtue of disorder averaging. However it does not explain the detailed structure of the chromatin regions studied. }
\end{figure}

Fig.~\ref{fig:exp:moments} shows the moment ratios which have been studied in Figs.~\ref{fig:ete:moments} and ~\ref{fig:ic:moments} together with the experimental data. The errorbars represent the propagated standard error of the measurements. They are quite large as we look on higher-order moments and the number of measurements per genomic distance is in the order of 50-100. Shown are also results from the RL model for a chain without excluded volume of length $N=1000$ and a looping probability of $p=7\times 10^{-5}$. Strikingly, in all three cases the moment ratios for the experimental data are larger than for a random walk, self-avoiding walk. This holds especially true for compact polymers where the moment ratios are even smaller. Note that for simplicity Fig.~\ref{fig:exp:moments} shows the ratios for the end-to-end distances, as we have shown in Fig.~\ref{fig:ic:moments}, the ratios for intrachain distances are even smaller. Obviously chromatin organization is not just a compacted state in terms of polymer physics, i.e. a chain in a poor solvent. The data suggests that there are structural features causing the fluctuations to be larger than even for the random walk model. The RL model offers an explanation for these experimental findings: Actually, chromatin-chromatin interactions are quite dynamic and there is a cell-to-cell variation not only in the thermal ensemble but also in the ensemble of different loop configuration. This disorder average gives rise to moment ratios extending  well beyond the random walk limit (see Fig.~\ref{fig:exp:moments}). However, the RL model still shows significant deviations from the experimental data. This is not surprising as the RL model as presented here assumes a homogeneous structure along the polymer while it is well-established that this is not the case~\cite{Shopland2006, Goetze2007}.
Nevertheless, this analysis clearly shows that the RL model extends beyond standard polymer models with attractive interactions and is in better agreement with experimental data on chromatin folding without assuming any a priori confinement of the chromosomes as has been done in other studies~\cite{Hahnfeldt1993}. 

\section{Conclusions}
In this study we have investigated the conformational and statistical properties of an ensemble of compact polymers. 

The algorithm we use here is basically the one proposed by Mansfield~\cite{Mansfield2006} for unbiased sampling Hamiltonian paths on a cubic $L\times L\times L$ lattice. As we are interested not only in Hamiltonian paths, but a broader class of compact polymers  we modified the algorithm by allowing vacancies on the cubic lattice such that we are able to sample compact polymers in the regime where $T\neq 0$ ---  or in terms of the lattice occupancy fraction $\rho\neq 1$. Using a highly parallel computing system we sampled conformations up to chain lengths of $N=256\,000$ and three different densities $\rho=0.1$, $\rho=0.5$ and $\rho=1$. 

While a lot of studies are devoted to the properties of compact polymers, little attention has been paid to the end-to-end distance distribution as well as to the distribution of intrachain segments within a globular polymer. However, this information is crucial for comparison to experimental data~\cite{Mateos2009}. We propose that the scaling function introduced by Fisher~\cite{Fisher1966} for the end-to-end distance distribution also holds approximately for compact polymers. We determine the exponents $\delta$ and $\mu$ (see eq.~\eqref{eq:ete:pr_analytic}) by a fit to the simulation data. One of the findings is that the scaling exponents do not depend on the density of the system studied. 
While the end-to-end distributions contain a very detailed information on the underlying polymer model, it is often sufficient to look at the first moments of this distribution. We analyze three dimensionless ratios of the end-to-end distribution function and show that deviations from a self-avoiding walk behavior become apparent from the fourth moment on when looking at the ratio $\left<R^4\right>/\left<R^2\right>$. 

From an experimental point of view it is often more interesting to know the distance distribution between two monomers along the contour of the chain. Therefore we analyze the distance distribution of the end points of segments of length $N_m$ inside a compact polymer of length $N$. We find that on very short contour lengths $N_m$ there is a random walk behavior (screening effect) while for larger contour lengths the moment ratios pretty fast fall below the ratios of the end-to-end distances. 

We also investigate the screening of excluded volume in compact polymers, which was expected from mean-field theory to exist not only in polymer melts, but also in compact polymers~\cite{Grosberg1994}. We find that there is a dominant crossover from the scaling $\left<R^2\right> \sim N^{2\nu}$ to $\left<R^2\right>\sim N$ with increasing density (Fig.~\ref{fig:screening:r2}). To estimate the screening length we carefully evaluate the structure function and find (Fig.~\ref{fig:screening:sq}) that the behavior is indeed similar to a polymer melt.

The aim of this study is not purely to enlighten us concerning the theoretical knowledge of compact polymers, but also to apply the findings to experimental data. One prime example where the results on distance distributions can be directly compared to is the organization of the chromatin fiber inside the human interphase nucleus. With fluorescent labeling experiments one can determine the distance between two markers separated by a certain genomic distance $g$, which is related to contour length $N_m$ in the simulations. We find that the chromatin fiber, although in a compact state that pretty much resembles a globular polymer behavior, shows significant deviations from these type of polymers. We propose that this is due to the dynamics of loop formation and unfolding, which is described well by the random loop model~\cite{Bohn2007a, Mateos2009}.

\begin{acknowledgments}
Monte Carlo simulations have been performed on the HELICS2 cluster at the Center for Scientific Computing (IWR) in Heidelberg. M.B. gratefully acknowledges funding from the Landesgraduiertenf\"orderung Baden-W\"urttemberg and support from the Heidelberg Graduate School of Mathematical and Computational Methods for the Sciences. 
\end{acknowledgments}

{\sffamily

}

\begin{thebibliography}{27}
\expandafter\ifx\csname natexlab\endcsname\relax\def\natexlab#1{#1}\fi
\expandafter\ifx\csname bibnamefont\endcsname\relax
  \def\bibnamefont#1{#1}\fi
\expandafter\ifx\csname bibfnamefont\endcsname\relax
  \def\bibfnamefont#1{#1}\fi
\expandafter\ifx\csname citenamefont\endcsname\relax
  \def\citenamefont#1{#1}\fi
\expandafter\ifx\csname url\endcsname\relax
  \def\url#1{\texttt{#1}}\fi
\expandafter\ifx\csname urlprefix\endcsname\relax\def\urlprefix{URL }\fi
\providecommand{\bibinfo}[2]{#2}
\providecommand{\eprint}[2][]{\url{#2}}

\bibitem[{\citenamefont{Grosberg and Khokhlov}(1994)}]{Grosberg1994}
\bibinfo{author}{\bibfnamefont{A.~Y.} \bibnamefont{Grosberg}} \bibnamefont{and}
  \bibinfo{author}{\bibfnamefont{A.~R.} \bibnamefont{Khokhlov}},
  \emph{\bibinfo{title}{Statistical Physics of Macromolecules}}
  (\bibinfo{publisher}{AIP Press}, \bibinfo{year}{1994}).

\bibitem[{\citenamefont{Dill et~al.}(1995)\citenamefont{Dill, Bromberg, Yue,
  Fiebig, Yee, Thomas, and Chan}}]{Dill1995}
\bibinfo{author}{\bibfnamefont{K.~A.} \bibnamefont{Dill}},
  \bibinfo{author}{\bibfnamefont{S.}~\bibnamefont{Bromberg}},
  \bibinfo{author}{\bibfnamefont{K.}~\bibnamefont{Yue}},
  \bibinfo{author}{\bibfnamefont{K.~M.} \bibnamefont{Fiebig}},
  \bibinfo{author}{\bibfnamefont{D.~P.} \bibnamefont{Yee}},
  \bibinfo{author}{\bibfnamefont{P.~D.} \bibnamefont{Thomas}},
  \bibnamefont{and} \bibinfo{author}{\bibfnamefont{H.~S.} \bibnamefont{Chan}},
  \bibinfo{journal}{Protein Sci} \textbf{\bibinfo{volume}{4}},
  \bibinfo{pages}{561} (\bibinfo{year}{1995}).

\bibitem[{\citenamefont{Lau and Dill}(1989)}]{Lau1989}
\bibinfo{author}{\bibfnamefont{K.~F.} \bibnamefont{Lau}} \bibnamefont{and}
  \bibinfo{author}{\bibfnamefont{K.~A.} \bibnamefont{Dill}},
  \bibinfo{journal}{Macromolecules} \textbf{\bibinfo{volume}{22}},
  \bibinfo{pages}{3986} (\bibinfo{year}{1989}).

\bibitem[{\citenamefont{Shakhnovich and Gutin}(1993)}]{Shakhnovich1993}
\bibinfo{author}{\bibfnamefont{E.}~\bibnamefont{Shakhnovich}} \bibnamefont{and}
  \bibinfo{author}{\bibfnamefont{A.}~\bibnamefont{Gutin}},
  \bibinfo{journal}{Proc. Natl. Acad. Sci. U. S. A.}
  \textbf{\bibinfo{volume}{90}}, \bibinfo{pages}{7195} (\bibinfo{year}{1993}),
  ISSN \bibinfo{issn}{{0027-8424}}.

\bibitem[{\citenamefont{M\"unkel and Langowski}(1998)}]{Munkel1998}
\bibinfo{author}{\bibfnamefont{C.}~\bibnamefont{M\"unkel}} \bibnamefont{and}
  \bibinfo{author}{\bibfnamefont{J.}~\bibnamefont{Langowski}},
  \bibinfo{journal}{Phys. Rev. E} \textbf{\bibinfo{volume}{57}},
  \bibinfo{pages}{5888} (\bibinfo{year}{1998}).

\bibitem[{\citenamefont{Goetze et~al.}(2007)\citenamefont{Goetze,
  Mateos-Langerak, Gierman, de~Leeuw, Giromus, Indemans, Koster, Ondrej,
  Versteeg, and van Driel}}]{Goetze2007}
\bibinfo{author}{\bibfnamefont{S.}~\bibnamefont{Goetze}},
  \bibinfo{author}{\bibfnamefont{J.}~\bibnamefont{Mateos-Langerak}},
  \bibinfo{author}{\bibfnamefont{H.~J.} \bibnamefont{Gierman}},
  \bibinfo{author}{\bibfnamefont{W.}~\bibnamefont{de~Leeuw}},
  \bibinfo{author}{\bibfnamefont{O.}~\bibnamefont{Giromus}},
  \bibinfo{author}{\bibfnamefont{M.~H.~G.} \bibnamefont{Indemans}},
  \bibinfo{author}{\bibfnamefont{J.}~\bibnamefont{Koster}},
  \bibinfo{author}{\bibfnamefont{V.}~\bibnamefont{Ondrej}},
  \bibinfo{author}{\bibfnamefont{R.}~\bibnamefont{Versteeg}}, \bibnamefont{and}
  \bibinfo{author}{\bibfnamefont{R.}~\bibnamefont{van Driel}},
  \bibinfo{journal}{Mol. Cell. Biol.} \textbf{\bibinfo{volume}{27}},
  \bibinfo{pages}{4475} (\bibinfo{year}{2007}).

\bibitem[{\citenamefont{Mateos-Langerak
  et~al.}(2009)\citenamefont{Mateos-Langerak, Bohn, de~Leeuw, Giromus, Manders,
  Verschure, Indemans, Gierman, Heermann, van Driel et~al.}}]{Mateos2009}
\bibinfo{author}{\bibfnamefont{J.}~\bibnamefont{Mateos-Langerak}},
  \bibinfo{author}{\bibfnamefont{M.}~\bibnamefont{Bohn}},
  \bibinfo{author}{\bibfnamefont{W.}~\bibnamefont{de~Leeuw}},
  \bibinfo{author}{\bibfnamefont{O.}~\bibnamefont{Giromus}},
  \bibinfo{author}{\bibfnamefont{E.~M.~M.} \bibnamefont{Manders}},
  \bibinfo{author}{\bibfnamefont{P.~J.} \bibnamefont{Verschure}},
  \bibinfo{author}{\bibfnamefont{M.~H.~G.} \bibnamefont{Indemans}},
  \bibinfo{author}{\bibfnamefont{H.~J.} \bibnamefont{Gierman}},
  \bibinfo{author}{\bibfnamefont{D.~W.} \bibnamefont{Heermann}},
  \bibinfo{author}{\bibfnamefont{R.}~\bibnamefont{van Driel}} and
  \bibinfo{author}{\bibfnamefont{S.}~\bibnamefont{Goetze}},
  \bibinfo{journal}{Proceedings of the National Academy
  of Sciences} \textbf{\bibinfo{volume}106}, \bibinfo{pages}{3812} (\bibinfo{year}{2009}). 

\bibitem[{\citenamefont{Eizenberg and Klafter}(1993)}]{Eizenberg1993}
\bibinfo{author}{\bibfnamefont{N.}~\bibnamefont{Eizenberg}} \bibnamefont{and}
  \bibinfo{author}{\bibfnamefont{J.}~\bibnamefont{Klafter}},
  \bibinfo{journal}{The Journal of Chemical Physics}
  \textbf{\bibinfo{volume}{99}}, \bibinfo{pages}{3976} (\bibinfo{year}{1993}).

\bibitem[{\citenamefont{Madras and Sokal}(1988)}]{Madras1988}
\bibinfo{author}{\bibfnamefont{N.}~\bibnamefont{Madras}} \bibnamefont{and}
  \bibinfo{author}{\bibfnamefont{A.}~\bibnamefont{Sokal}}, \bibinfo{journal}{J.
  Stat. Phys.} \textbf{\bibinfo{volume}{50}}, \bibinfo{pages}{109}
  (\bibinfo{year}{1988}).

\bibitem[{\citenamefont{Rapaport}(1985)}]{Rapaport1985}
\bibinfo{author}{\bibfnamefont{D.~C.} \bibnamefont{Rapaport}},
  \bibinfo{journal}{Journal of Physics A: Mathematical and General}
  \textbf{\bibinfo{volume}{18}}, \bibinfo{pages}{113} (\bibinfo{year}{1985}).

\bibitem[{\citenamefont{Domb and Hioe}(1969)}]{Domb1969}
\bibinfo{author}{\bibfnamefont{C.}~\bibnamefont{Domb}} \bibnamefont{and}
  \bibinfo{author}{\bibfnamefont{F.~T.} \bibnamefont{Hioe}},
  \bibinfo{journal}{The Journal of Chemical Physics}
  \textbf{\bibinfo{volume}{51}}, \bibinfo{pages}{1915} (\bibinfo{year}{1969}).

\bibitem[{\citenamefont{Shakhnovich and Gutin}(1990)}]{Shakhnovich1990}
\bibinfo{author}{\bibfnamefont{E.}~\bibnamefont{Shakhnovich}} \bibnamefont{and}
  \bibinfo{author}{\bibfnamefont{A.}~\bibnamefont{Gutin}}, \bibinfo{journal}{J.
  Chem. Phys.} \textbf{\bibinfo{volume}{93}}, \bibinfo{pages}{5967}
  (\bibinfo{year}{1990}).

\bibitem[{\citenamefont{{Pande} et~al.}(1996)\citenamefont{{Pande}, {Joerg},
  {Grosberg}, and {Tanaka}}}]{Pande1996}
\bibinfo{author}{\bibfnamefont{V.~S.} \bibnamefont{{Pande}}},
  \bibinfo{author}{\bibfnamefont{C.}~\bibnamefont{{Joerg}}},
  \bibinfo{author}{\bibfnamefont{A.~Y.} \bibnamefont{{Grosberg}}},
  \bibnamefont{and} \bibinfo{author}{\bibfnamefont{T.}~\bibnamefont{{Tanaka}}},
  \bibinfo{journal}{Journal of Physics A Mathematical General}
  \textbf{\bibinfo{volume}{29}}, \bibinfo{pages}{4753} (\bibinfo{year}{1996}).

\bibitem[{\citenamefont{Ramakrishnan et~al.}(1995)\citenamefont{Ramakrishnan,
  Pekny, and Caruthers}}]{Ramakrishnan1995}
\bibinfo{author}{\bibfnamefont{R.}~\bibnamefont{Ramakrishnan}},
  \bibinfo{author}{\bibfnamefont{J.~F.} \bibnamefont{Pekny}}, \bibnamefont{and}
  \bibinfo{author}{\bibfnamefont{J.~M.} \bibnamefont{Caruthers}},
  \bibinfo{journal}{J. Chem. Phys.} \textbf{\bibinfo{volume}{103}},
  \bibinfo{pages}{7592} (\bibinfo{year}{1995}).

\bibitem[{\citenamefont{Mansfield}(2006)}]{Mansfield2006}
\bibinfo{author}{\bibfnamefont{M.~L.} \bibnamefont{Mansfield}},
  \bibinfo{journal}{J. Chem. Phys.} \textbf{\bibinfo{volume}{125}},
  \bibinfo{eid}{154103} (pages~\bibinfo{numpages}{7}) (\bibinfo{year}{2006}).

\bibitem[{\citenamefont{Lua et~al.}(2004)\citenamefont{Lua, Borovinskiy, and
  Grosberg}}]{Lua2004}
\bibinfo{author}{\bibfnamefont{R.}~\bibnamefont{Lua}},
  \bibinfo{author}{\bibfnamefont{A.~L.} \bibnamefont{Borovinskiy}},
  \bibnamefont{and} \bibinfo{author}{\bibfnamefont{A.~Y.}
  \bibnamefont{Grosberg}}, \bibinfo{journal}{Polymer}
  \textbf{\bibinfo{volume}{45}}, \bibinfo{pages}{717} (\bibinfo{year}{2004}).

\bibitem[{\citenamefont{Bohn et~al.}(2007)\citenamefont{Bohn, Heermann, and van
  Driel}}]{Bohn2007a}
\bibinfo{author}{\bibfnamefont{M.}~\bibnamefont{Bohn}},
  \bibinfo{author}{\bibfnamefont{D.~W.} \bibnamefont{Heermann}},
  \bibnamefont{and} \bibinfo{author}{\bibfnamefont{R.}~\bibnamefont{van
  Driel}}, \bibinfo{journal}{Physical Review E (Statistical, Nonlinear, and
  Soft Matter Physics)} \textbf{\bibinfo{volume}{76}}, \bibinfo{eid}{051805}
  (pages~\bibinfo{numpages}{8}) (\bibinfo{year}{2007}).

\bibitem[{\citenamefont{Binder and Heermann}(2002)}]{Binder2002}
\bibinfo{author}{\bibfnamefont{K.}~\bibnamefont{Binder}} \bibnamefont{and}
  \bibinfo{author}{\bibfnamefont{D.}~\bibnamefont{Heermann}},
  \emph{\bibinfo{title}{Monte Carlo Simulation in Statistical Physics. An
  Introduction.}} (\bibinfo{publisher}{Springer}, \bibinfo{year}{2002}),
  \bibinfo{edition}{4th} ed.

\bibitem[{\citenamefont{Le~Guillou and Zinn-Justin}(1980)}]{Le1980}
\bibinfo{author}{\bibfnamefont{J.~C.} \bibnamefont{Le~Guillou}}
  \bibnamefont{and}
  \bibinfo{author}{\bibfnamefont{J.}~\bibnamefont{Zinn-Justin}},
  \bibinfo{journal}{Phys. Rev. B} \textbf{\bibinfo{volume}{21}},
  \bibinfo{pages}{3976} (\bibinfo{year}{1980}).

\bibitem[{\citenamefont{Fisher}(1966)}]{Fisher1966}
\bibinfo{author}{\bibfnamefont{M.}~\bibnamefont{Fisher}}, \bibinfo{journal}{J.
  Chem. Phys.} \textbf{\bibinfo{volume}{44}}, \bibinfo{pages}{616}
  (\bibinfo{year}{1966}).

\bibitem[{\citenamefont{de~Gennes}(1979)}]{Gennes1979}
\bibinfo{author}{\bibfnamefont{P.-G.} \bibnamefont{de~Gennes}},
  \emph{\bibinfo{title}{Scaling concepts in polymer physics}}
  (\bibinfo{publisher}{Ithaca, N.Y., Cornell University Press},
  \bibinfo{year}{1979}).

\bibitem[{\citenamefont{Paul et~al.}(1991)\citenamefont{Paul, Binder, Heermann,
  and Kremer}}]{Paul1991}
\bibinfo{author}{\bibfnamefont{W.}~\bibnamefont{Paul}},
  \bibinfo{author}{\bibfnamefont{K.}~\bibnamefont{Binder}},
  \bibinfo{author}{\bibfnamefont{D.~W.} \bibnamefont{Heermann}},
  \bibnamefont{and} \bibinfo{author}{\bibfnamefont{K.}~\bibnamefont{Kremer}},
  \bibinfo{journal}{Journal De Physique Ii} \textbf{\bibinfo{volume}{1}},
  \bibinfo{pages}{37} (\bibinfo{year}{1991}).

\bibitem[{\citenamefont{Bruns}(1992)}]{Bruns1992}
\bibinfo{author}{\bibfnamefont{W.}~\bibnamefont{Bruns}},
  \bibinfo{journal}{Makromolekulare Chemie-Theory And Simulations}
  \textbf{\bibinfo{volume}{1}}, \bibinfo{pages}{287} (\bibinfo{year}{1992}).

\bibitem[{\citenamefont{\v{S}olc and Stockmayer}(1971)}]{Solc1971}
\bibinfo{author}{\bibfnamefont{K.}~\bibnamefont{\v{S}olc}} \bibnamefont{and}
  \bibinfo{author}{\bibfnamefont{W.~H.} \bibnamefont{Stockmayer}},
  \bibinfo{journal}{The Journal of Chemical Physics}
  \textbf{\bibinfo{volume}{54}}, \bibinfo{pages}{2756} (\bibinfo{year}{1971}).

\bibitem[{\citenamefont{Schiessel et~al.}(2001)\citenamefont{Schiessel,
  Gelbart, and Bruinsma}}]{Schiessel2001}
\bibinfo{author}{\bibfnamefont{H.}~\bibnamefont{Schiessel}},
  \bibinfo{author}{\bibfnamefont{W.~M.} \bibnamefont{Gelbart}},
  \bibnamefont{and} \bibinfo{author}{\bibfnamefont{R.}~\bibnamefont{Bruinsma}},
  \bibinfo{journal}{Biophys. J.} \textbf{\bibinfo{volume}{80}},
  \bibinfo{pages}{1940} (\bibinfo{year}{2001}).

\bibitem[{\citenamefont{Shopland et~al.}(2006)\citenamefont{Shopland, Lynch,
  Peterson, Thornton, Kepper, Hase, Stein, Vincent, Molloy, Kreth
  et~al.}}]{Shopland2006}
\bibinfo{author}{\bibfnamefont{L.~S.} \bibnamefont{Shopland}},
  \bibinfo{author}{\bibfnamefont{C.~R.} \bibnamefont{Lynch}},
  \bibinfo{author}{\bibfnamefont{K.~A.} \bibnamefont{Peterson}},
  \bibinfo{author}{\bibfnamefont{K.}~\bibnamefont{Thornton}},
  \bibinfo{author}{\bibfnamefont{N.}~\bibnamefont{Kepper}},
  \bibinfo{author}{\bibfnamefont{J.~v.} \bibnamefont{Hase}},
  \bibinfo{author}{\bibfnamefont{S.}~\bibnamefont{Stein}},
  \bibinfo{author}{\bibfnamefont{S.}~\bibnamefont{Vincent}},
  \bibinfo{author}{\bibfnamefont{K.~R.} \bibnamefont{Molloy}},
  \bibinfo{author}{\bibfnamefont{G.}~\bibnamefont{Kreth}},
  \bibnamefont{et~al.}, \bibinfo{journal}{J. Cell Biol.}
  \textbf{\bibinfo{volume}{174}}, \bibinfo{pages}{27} (\bibinfo{year}{2006}).

\bibitem[{\citenamefont{Hahnfeldt et~al.}(1993)\citenamefont{Hahnfeldt, Hearst,
  Brenner, Sachs, and Hlatky}}]{Hahnfeldt1993}
\bibinfo{author}{\bibfnamefont{P.}~\bibnamefont{Hahnfeldt}},
  \bibinfo{author}{\bibfnamefont{J.~E.} \bibnamefont{Hearst}},
  \bibinfo{author}{\bibfnamefont{D.~J.} \bibnamefont{Brenner}},
  \bibinfo{author}{\bibfnamefont{R.~K.} \bibnamefont{Sachs}}, \bibnamefont{and}
  \bibinfo{author}{\bibfnamefont{L.~R.} \bibnamefont{Hlatky}},
  \bibinfo{journal}{Proc. Natl. Acad. Sci. U. S. A.}
  \textbf{\bibinfo{volume}{90}}, \bibinfo{pages}{7854} (\bibinfo{year}{1993}).

\end{thebibliography}
\end{document}